\documentclass[printer]{aa}
\usepackage{graphicx}
\usepackage{hyperref}

\usepackage{natbib}
\bibpunct{(}{)}{;}{a}{}{,} 

\usepackage{txfonts}
\usepackage{longtable}
\usepackage{listings}
\usepackage{color}
\usepackage{textcomp}

\begin{document}

\title{
A \textit{Gaia}~DR2 view of the Open Cluster population in the Milky Way
}

   \subtitle{}

\author{
T. Cantat-Gaudin\inst{\ref{IEECUB}}
\and
C. Jordi\inst{\ref{IEECUB}}
\and
A. Vallenari\inst{\ref{OAPD}}
\and
A. Bragaglia\inst{\ref{OABO}}
\and
L. Balaguer-N{\'u}{\~n}ez\inst{\ref{IEECUB}}	
\and
C. Soubiran\inst{\ref{LAB}}
\and
D. Bossini\inst{\ref{OAPD}}
\and
A. Moitinho\inst{\ref{SIMUL}}
\and
A. Castro-Ginard\inst{\ref{IEECUB}}
\and
A. Krone-Martins\inst{\ref{SIMUL}}
\and
L. Casamiquela\inst{\ref{LAB}}
\and
R. Sordo\inst{\ref{OAPD}}
\and
R. Carrera\inst{\ref{OAPD}}
}

\institute{
Institut de Ci\`encies del Cosmos, Universitat de Barcelona (IEEC-UB), Mart\'i i Franqu\`es 1, E-08028 Barcelona, Spain\label{IEECUB}
\and
INAF-Osservatorio Astronomico di Padova, vicolo Osservatorio 5, 35122 Padova, Italy\label{OAPD}
\and
INAF-Osservatorio di Astrofisica e Scienza dello Spazio, via Gobetti 93/3, 40129 Bologna, Italy\label{OABO}
\and
Laboratoire d’Astrophysique de Bordeaux, Univ. Bordeaux, CNRS, UMR 5804, 33615 Pessac, France\label{LAB}
\and
CENTRA, Faculdade de Ci\^encias, Universidade de Lisboa, Ed. C8, Campo Grande, P-1749-016 Lisboa, Portugal\label{SIMUL}
}

\date{Received date / Accepted date }

\abstract
{Open clusters are convenient probes of the structure and history of the Galactic disk. They are also fundamental to stellar evolution studies. The second \textit{Gaia} data release contains precise astrometry at the sub-milliarcsecond level and homogeneous photometry at the mmag level, that can be used to characterise a large number of clusters over the entire sky.}
{In this study we aim to establish a list of members and derive mean parameters, in particular distances, for as many clusters as possible, making use of \textit{Gaia} data alone.}
{We compile a list of thousands of known or putative clusters from the literature. We then apply an unsupervised membership assignment code, UPMASK, to the \textit{Gaia}~DR2 data contained within the fields of those clusters.}
{We obtained a list of members and cluster parameters for 1229 clusters. As expected, the youngest clusters are seen to be tightly distributed near the Galactic plane and to trace the spiral arms of the Milky Way, while older objects are more uniformly distributed, deviate further from the plane, and tend to be located at larger Galactocentric distances. Thanks to the quality of \textit{Gaia}~DR2 astrometry, the fully homogeneous parameters derived in this study are the most precise to date. Furthermore, we report on the serendipitous discovery of 60 new open clusters in the fields analysed during this study.}
{}


\maketitle{}

\section{Introduction}
Our vantage point inside the disk of the Milky Way allows us to see in great detail some of the finer structures present in the solar neighbourhood, but impedes our understanding of the three-dimensional structure of the disk on a larger scale. In order to reconstruct the overall shape of our Galaxy, it is necessary to estimate distances to astronomical objects that we use as tracers, and study their distribution. Since the historical works of \citet{Herschel1785}, who estimated photometric distances to field stars, a variety of tracers have been used, such as planetary nebulae, RR Lyrae, Cepheids, OB stars, or HII regions. An abundant literature focuses on clusters as tracers of the Galactic disk.

The stellar clusters belonging to the disk of the Galaxy are traditionally refered to as open clusters (OCs). As simple stellar populations, their ages and distances can be estimated in a relatively simple (albeit model-dependent) way by means of photometry, making them convenient tracers of the structure of the Milky Way \citep[see e.g.][]{Janes82,Dias05,Piskunov06,Moitinho10,Buckner14}. They have been used as such since the study of \citet{Trumpler30}, proving the existence of absorption by the interstellar medium.
They are also popular tracers to follow the metallicity gradient of the Milky Way \citep[a non-exhaustive list includes][]{Janes79,Friel95,Twarog97,Yong05,Bragaglia06,Carrera11,Netopil16,Casamiquela17} and its evolution through time \citep[e.g.][]{Friel02,Magrini09,Yong12,Frinchaboy13,Jacobson16}, providing insight on the formation of the Galactic disk. Studies of the kinematics of OCs and reconstructions of their individual orbits \citep{Wu09,VandePutte10,CantatGaudin16,Reddy16} help us understand the internal processes of heating \citep{MartinezMedina16,Gustafsson16,Quillen18} and radial migration \citep{Roskar08,Minchev16,Anders17}, and how they affect the chemodynamical evolution of the disk. Some very perturbed orbits might also provide evidence for recent merger events and traces of past accretion from outside the Galaxy \citep{Law10,CantatGaudin16}.

Open clusters are not only useful tracers of the Milky Way structure but are also interesting targets in their own right. They are homogeneous groups of stars with the same age and same initial chemical composition, formed in a single event from the same gas cloud, and therefore constitute ideal laboratories to study stellar formation and evolution. Although most stars in the Milky Way are observed in isolation, it is believed that most (possibly all) stars form in clustered environments and spend at least a short amount of time gravitationally bound with their siblings \citep[see e.g.][]{Clarke2000,Lada03,PortegiesZwart10}, embedded in their progenitor molecular cloud. A majority of such systems will be disrupted in their first few million years of existence, due to mechanisms possibly involving gas loss driven by stellar feedback \citep{Moeckel10,Brinkmann17} or encounters with giant molecular clouds \citep{Gieles06}. Nonetheless, a fraction will survive the embedded phase and remain bound over longer timescales. 

Some of the most popular catalogues gathering information on OCs in the Milky Way include the WEBDA database \citep{Mermilliod95}, and the catalogues of \citet[][hereafter DAML]{Dias02} and \citet[][hereafter MWSC]{Kharchenko13}. 
The latest recent version of the DAML catalogue lists about 2200 objects, most of them located within 2\,kpc of the Sun, while MWSC lists over 3000 objects (including globular clusters), many of which are putative clusters needing confirmation. 

Although claims have been made that the sample of known OCs might be complete out to distances of 1.8\,kpc \citep{Kharchenko13,Joshi16,Yen18}, it is likely that some objects are still left to be found in the solar neighbourhood, in particular old OCs, as pointed out by \citet{Moitinho10} and \citet{Piskunov18}.
Sparse nearby OCs with large apparent sizes that do not stand out as significant overdensities in the sky can also be revealed by the use of astrometric data, as the recent discoveries of \citet{Roeser16} and \citet{Castro18} have shown.

The inhomogeneous analysis of the cluster population often lead to discrepant values, due to the use of different data and methods of analysis. This was noted for instance by \citet{Dias14} and \citet{Netopil15}. Characterising OCs is often done with the use of data of different nature, combining photometry from dedicated observations such as the Bologna Open Clusters Chemical Evolution project \citep{Bragaglia06}, the WIYN Open Cluster Study \citep{AnthonyTwarog16} or the Open Cluster Chemical Abundances from Spanish Observatories program \citep{Casamiquela16}.
Other studies make use of data from all-sky surveys \citep[2MASS][is a popular choice for studies inside the Galactic plane]{Skrutskie06}, proper motions from the all-sky catalogues Tycho-2 \citep{Hog00}, PPMXL \citep{Roeser10}, or UCAC4 \citep{Zacharias13}, or parallaxes from \textsc{Hipparcos} \citep{ESA97,Perryman97,vanLeeuwen07}. The study of \citet{Sampedro17} reports membership for 1876 clusters, based on UCAC4 proper motions alone.
The ongoing ESA mission \textit{Gaia} \citep{Perryman01,Gaia16prusti} is carrying out an unprecedented astrometric, photometric, and spectroscopic all-sky survey, reducing the need for cross-matching catalogues or compiling complementary data.

Space-based astrometry in all-sky surveys has enabled membership determinations from a full astrometric solution (using proper motions and parallaxes), such as the studies of \citet{Robichon99} (50 OCs with at least 4 stars, within 500\,pc) and \citet{vanLeeuwen09} (20 OCs) using \textsc{Hipparcos} data, or \citet{FvL17} (19 OCs within 500\,pc) using the Tycho-\textit{Gaia} Astrometric Solution \citep[TGAS,][]{Michalik15,GaiaDR1}. \citet{Yen18} have determined membership for stars in 24 OCs, adding fainter members, for clusters within 333\,pc. 

The study of \citet{CantatGaudin18} established membership for 128 OCs based on the proper motions and parallaxes of TGAS, complementing the TGAS proper motions with UCAC4 data \citep{Zacharias13} and 2MASS photometry \citep{Skrutskie06}. The catalogue of the second \textit{Gaia} data release \citep[][hereafter \textit{Gaia}~DR2]{GDR2content} reaches a $G$-band magnitude of 21 (9 magnitudes fainter than TGAS). At its faint end, the \textit{Gaia}~DR2 astrometric precision is comparable with that of TGAS, while for stars brighter than $G\sim15$ the precision is about ten times better than in TGAS, allowing us to extend membership determinations to fainter stars and to characterise more distant objects. The \textit{Gaia}~DR2 catalogue also contains magnitudes in the three passbands of the \textit{Gaia} photometric system $G$, $G_{BP}$, $G_{RP}$ (where TGAS only featured $G$-band magnitudes) with precisions at the mmag level. One of the most precious information provided with \textit{Gaia}~DR2 are individual parallaxes to more than a billion stars, from which distances can be inferred for a large number of clusters.

This paper aims to provide a view of the Milky Way cluster population by establishing a list of cluster members through the use of \textit{Gaia}~DR2 data only. It is organised as follows: 
Section~\ref{sec:data} presents the \textit{Gaia}~DR2 data used in this study.
Section~\ref{sec:method} describes our tools and approach to membership selection.
Section~\ref{sec:meanparams} presents the individual parameters and distances derived for the detected clusters, and Sect.~\ref{sec:remarks} comments on some specific objects.
Section~\ref{sec:galaxy} places the clusters in the context of the Galactic disk.
Section~\ref{sec:discussion} contains a discussion, and Sect.~\ref{sec:conclusion} closing remarks.

\section{The data} \label{sec:data}

\subsection{The multi-dimensional dataset of \textit{Gaia}~DR2}
\begin{figure}[ht]
\begin{center} \resizebox{\hsize}{!}{\includegraphics[scale=0.5]{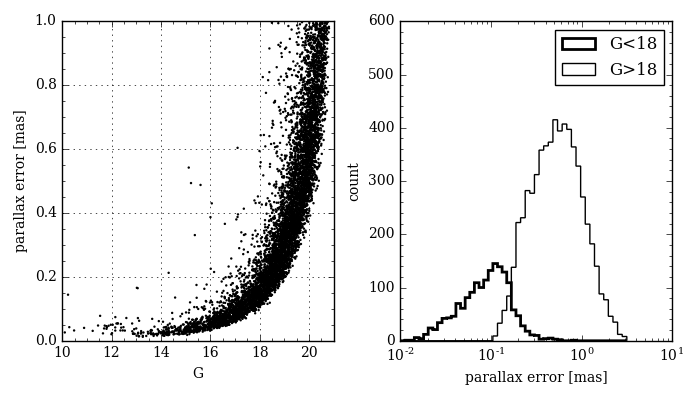}} \caption{\label{fig:errors_G18} Left: parallax nominal error agains $G$ magnitude for a random sample of 10\,000 \textit{Gaia}~DR2 stars. Right: distribution of nominal parallax error for the stars brighter than $G=18$ (thick histogram) and fainter (thin histogram) of the same random sample. } \end{center}
\end{figure}

The 1.7-billion-source catalogue of \textit{Gaia}~DR2 is unprecedented for an astronomical dataset in terms of its sheer size, high-dimensionality, and astrometric precision and accuracy. In particular, it provides a 5-parameter astrometric solution (proper motions in right ascension and declination $\mu_{\alpha*}$ and $\mu_{\delta}$ and parallaxes $\varpi$) and magnitudes in three photometric filters ($G$, $G_{BP}$, $G_{RP}$) for more than 1.3 billion sources \citep{GDR2content}. 
The large magnitude range it covers however leads to significant differences in precision between the bright and faint sources. At the bright end ($G<14$), the nominal uncertainties reach precisions of 0.02\,mas in parallax and 0.05\,mas\,yr$^{-1}$ in proper motions, while for sources near $G\sim21$ the uncertainties reach 2\,mas and 5\,mas\,yr$^{-1}$, respectively (see Fig.~\ref{fig:errors_G18}). In this study we only made use of sources brighter than $G=18$, corresponding to typical astrometric uncertainties of 0.3\,mas\,yr$^{-1}$ in proper motion and 0.15\,mas in parallax. In most open clusters, the contrast between cluster and field stars is very low for sources fainter than this limit (although in some cases such as Kronberger~31 or Saurer~1 hints of overdensities are visible in positional space), and their large proper motion and parallax uncertainties do not allow them to be seen as overdensities in astrometric space either. It should be possible to identify cluster members among the stars with large astrometric uncertainties (or among those for which \textit{Gaia}~DR2 does not provide an astrometric solution at all) if other criteria such as photometric selections are employed, although many of the sources without a full astrometric solution also lack $G_{BP}$ and $G_{RP}$ photometry.

In addition to discarding the least informative sources, this cut off value greatly dminishes to volume of data to process and makes computations faster, as 80\% of the \textit{Gaia}~DR2 sources are fainter than G$\sim$18. 
In terms of distances, this cut corresponds to the magnitude of the turn off stars in a 100\,Myr cluster ($G_{\mathrm{abs}}\sim-1.5$) seen at 80\,kpc, or in a 3\,Gyr cluster ($G_{\mathrm{abs}}\sim3$) seen at 10\,kpc (without considering interstellar extinction). We therefore expect the most distant and oldest known OCs to be near our detection threshold.

\begin{figure}[ht]
\begin{center} \resizebox{\hsize}{!}{\includegraphics[scale=0.5]{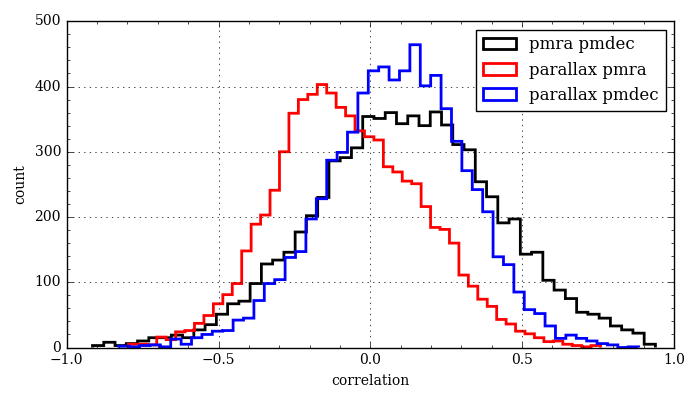}} \caption{\label{fig:correlations_hist_GDR2} Correlation coefficients between three astrometric parameters for a random sample of 10\,000 \textit{Gaia}~DR2 sources. } \end{center}
\end{figure}

The \textit{Gaia} astrometric solution is a simultaneous determination of the five parameters $(\alpha,\delta,\mu_{\alpha*},\mu_{\delta},\varpi)$, and the uncertainties on these five quantities present non-zero correlations, albeit not as strong as in TGAS. The importance of taking correlations into account when considering whether two data points are compatible within their uncertainties is shown in \citet{CantatGaudin18}. The correlation coefficients for a random sample of 10\,000 \textit{Gaia}~DR2 sources are shown in Fig.~\ref{fig:correlations_hist_GDR2}. 

Our membership assignment relies on the astrometric solution, and we only used the \textit{Gaia}~DR2 photometry to manually confirm that the groups identified matched the expected aspect of a cluster in a colour-magnitude diagram. We did not attempt to correct the photometry of individual sources from interstellar extinction using the \textit{Gaia}~DR2 values of $A_\mathrm{G}$ and $E\mathrm{(BP-RP)}$, as their nominal uncertainties \citep[$\sim0.46$\,mag in $A_\mathrm{G}$,][]{Andrae18} do not allow to improve the aspect of a cluster sequence in a colour-magnitude diagram. 

The \textit{Gaia}~DR2 data also contains radial velocities for about 7 million stars (mostly brighter than $G\sim13$), which we did not exploit in this work, but can provide valuable information for a number of OCs.

We queried the data through the ESAC portal\footnote{https://gea.esac.esa.int/archive/}, and scripted most queries using the package \texttt{pygacs}\footnote{https://github.com/Johannes-Sahlmann/pygacs}.

\subsection{Choice of OCs to target}

We compiled a list of 3328 known clusters and candidates taken from the catalogues of \citet{Dias02} and \citet{Kharchenko13}, and the publications of \citet{Froebrich07}, \citet{Schmeja14}, \citet{Scholz15}, and \citet{Roeser16}. We excluded three well-studied clusters a priori: Collinder~285 (the Ursa Major moving group), Melotte~25 (the Hyades) and Melotte~111 (Coma Ber), because their large extension across the sky makes them difficult to identify. We excluded all the objects listed as globular clusters in \citet{Kharchenko13} and in the latest update of the catalogue of \citet{Harris96} from our list of targets.

Many of the cluster candidates listed in the literature are so-called ``infrared clusters''. Those were discovered through observations in the near-infrared passbands of the 2MASS survey, observing at wavelengths at which the interstellar medium is more transparent than in optical, and therefore allowing to see further into the Galactic disk. Although \textit{Gaia} is observing at optical wavelengths, its $G$-band limit is 5\,mag fainter than the $J$-band completeness limit of 2MASS (in our case 2\,mag fainter, since in this study we only use sources with $G<18$), which should make most of the clusters detected in the 2MASS data observable by \textit{Gaia} as long as the extinction is lower than $A_V\sim5$.

\section{The method} \label{sec:method}

In this work we applied the membership assignment code UPMASK \citep[Unsupervised Photometric Membership Assignment in Stellar Clusters,][]{KroneMartins14} on stellar fields centred on each known cluster or candidate.

\subsection{UPMASK}
The classification scheme of UPMASK is unsupervised, and relies on no physical assumption about stellar clusters, apart from the fact that its member stars must share common properties, and be more tightly distributed on the sky than a random distribution. Although the original implementation of the method was created to stellar, photometric data,
the approach was designed to be easily generalized to other quantities or sources, (astronomical and non-astronomical alike, as galaxies or cells). The method was successfully applied to the astrometric data of the Tycho-\textit{Gaia} Astrometric Solution in \citet{CantatGaudin18}.

We recall here the main steps of the process:

\begin{enumerate}
	\item Small groups of stars are identified in the 3-dimensional astrometric space ($\mu_{\alpha*}$, $\mu_{\delta}$, $\varpi$) through k-means clustering.
	\item We assess whether the distribution on the sky of each of these small groups is more concentrated than a random distribution and return a binary ``yes'' or ``no'' answer for each group (this is referred to as the ``veto'' step).
\end{enumerate}

In this implementation we perfom the veto step by comparing the total branch length of the minimum spanning tree \citep[see][for a historical review]{Graham85} connecting the stars with the expected branch length for a random uniform distribution covering the investigated field of view. The assumption that the field star distribution is uniform might create false positives in regions where the background density is shaped by differential extinction, such as star-forming regions and around young OB associations (which were not included in this study). Artefacts in the density distribution caused by the \textit{Gaia} scanning law were significantly improved from \textit{Gaia}-DR1 to \textit{Gaia}-DR2 \citep{Arenou17,Arenou18} and do not significantly affect the stars brighter than $G=18$, which is the magnitude limit we adopted in this study.

To turn the binary yes/no flag into a membership probability, we redraw new values of ($\mu_{\alpha*}$, $\mu_{\delta}$, $\varpi$) for each source based on its listed value and uncertainty (and the correlations between those three parameters), and perform the grouping and the veto steps again. After a certain number of redrawings, the final probability is the frequency with which a given star passes the veto.

\subsection{Workflow}

\subsubsection{Finding the signature of the cluster}
For every cluster (or candidate cluster) under investigation, we started with a cone search centred on the position listed in the literature. The numbers listed for the apparent size of a cluster can vary significantly from one catalogue to another. The radius used was twice the value \texttt{Diam} listed in DAML, or the value \texttt{r2} for clusters only listed in MWSC. The size of the field of view is not critical to UPMASK  \citep[][have shown that the effect on the contamination and completeness of UPMASK is small when doubling the cluster radius]{KroneMartins14}, but a cluster might be missed if the sample only contains the dense inner regions. The contrast between cluster and field stars in astrometric space will not be optimal if the radius used is inappropriately large. 

In addition, we performed a broad selection in parallax, keeping only stars with $\varpi$ within 0.5\,mas of the parallax expected from their distance (or 0.5\,mas around the range of expected parallax, for clusters with discrepant distances in the catalogues). We performed no prior proper motion selection.
We ran UPMASK with 5 redrawings on each of the investigated fields. {We found that this number is a good compromise, as it is sufficient to reveal whether a statistically clustered group of stars exists, while performing more redrawings would render the task of investigating over 3000 fields of view even more computationally intensive. The procedure of querying the \textit{Gaia} archive and running the algorithm were fully automated. We inspected and controlled the output of the assignment manually.

\subsubsection*{Where nothing was detected}
Where no significant hint of a cluster was found, we performed the procedure again dividing the search radius by two in order to provide a better contrast between cluster and field stars. An additional 66 clusters were detected using this smaller field of view. Furtermore, we noticed that the apparent sizes listed for the FSR clusters differ by sometimes an order of magnitude between the DAML and MWSC catalogue, with larger diameters in the latter catalogue. Increasing the size of the field of view to up to 12 times \texttt{Diam} enabled us to recover 24 additional FSR clusters\footnote{For FSR~0686 we find a median radius r$_{50}$=0.156$^{\circ}$, comparable with the MWSC radius of 0.17$^{\circ}$, but at odds with the DAML diameter of 1.3\,arcmin ($\sim$0.02$^{\circ}$)}. 

For the clusters with expected distances under 2\,kpc, we inspected the fields one last time working only with stars brighter than $G=15$, but this final attempt failed to detect any more clusters. The most sparse and nearby clusters (such as Platais~2 or Collinder~65) are usually not detected by our algorithm, and should be investigated with tailored astrometric and photomeric preselection. At the distant end, it is likely that some non-detected clusters have stars in \textit{Gaia}~DR2 that can be identified with an appropriate initial selection, and possibly the use of non-\textit{Gaia} data. We also failed to find trace of the cluster candidates reported by \citet{Schmeja14} and \citet{Scholz15}, which are discussed in Sect.~\ref{sec:galaxy}.

\subsubsection*{Where more than one OC were detected in the same field}

In some cases clusters overlap on the sky, leading to multiple detections. Most of the time, such clusters can however be clearly separated in astrometric space or in a colour-magnitude diagram. In those cases (such as the pairs NGC~7245/King~9 or NGC~2451A/NGC~2451B) we manually devised appropriate cuts in proper motion and/or parallax. 

\subsubsection*{Where unreported clusters were found}
Although this study only aimed at characterising the known OCs and is not optimised for cluster detection, we found dozens of groups with consistent proper motions and parallaxes and a confirmed cluster-like sequence in a colour-magnitude diagram, that to our best knowledge were so far unreported. Those 60 clusters are discussed in Sect.~\ref{sec:gulliver}, and their positions and mean parameters are reported in Table~\ref{table:meanparams}.

\subsubsection{Running the algorithm on a restricted sample}
Once a centroid in ($\mu_{\alpha*}$, $\mu_{\delta}$, $\varpi$) was identified for all feasible OCs, we only selected stars with proper motions within 2\,mas\,yr$^{-1}$ of the identified overdensity, and parallaxes with 0.3\,mas. Those values were adopted because they allow to eliminate a large number of non-member stars, while being still larger than the apparent dispersion of the cluster members.
For a handful of nearby clusters with large apparent proper motion dispersions (Blanco~1, Mamajek~1, Melotte~20, Melotte~22, NGC~2451A, NGC~2451B, NGC~2632, Platais~3, Platais~8, Platais~9, and Platais~10) we did not restrict the proper motions. All of them are closer than 240\,pc except NGC~2451B for which we derive in this study a distance of 364\,pc. In the cases where the clusters present a very compact aspect in proper motion space (all more distant than 900\,pc), we selected sources with proper motions within 0.5\,mas\,yr$^{-1}$ of the centroid.

We then ran 10 iterations of UPMASK, in order to obtain membership probabilities from 0 to 100\% by increment of 10\%. We ended up with a set of 1229 clusters for which at least five stars have a membership probability greater than 50\% (the number of clusters for various combinations of threshold numbers and probabilities is shown in Table~\ref{table:NP}). Examples are shown in Fig.~\ref{fig:example_Berkeley_18} to Fig.~\ref{fig:example_Tombaugh_2}. The full membership list for all clusters is available as an electronic table.

\section{Astrometric parameters} \label{sec:meanparams}

\subsection{Main cluster parameters}
We computed the median $\mu_{\alpha*}$, $\mu_{\delta}$, and $\varpi$ of the probable cluster members (those with probabilities $>50$\%), after removing outliers discrepant from the median value by more than three median absolute deviations.
The values are reported in Table~\ref{table:meanparams}.

\begin{table*}
\begin{center}
	\caption{ \label{table:meanparams} Summary of mean parameters for the OCs characterised in this study (the full table is available in the electronic version of this paper).}
	\small\addtolength{\tabcolsep}{-2pt}
	\begin{tabular}{ c  c  c  c  c  c  c  c  c  c  c  c  c  c  c}
	\hline
	\hline

OC & $\alpha$ & $\delta$ & r$_{50}$ & $N$ & $\mu_{\alpha*}$ & $\sigma_{\mu_{\alpha*}}$ & $\mu_{\delta}$ & $\sigma_{\mu_{\delta}}$ & $\varpi$ & $\sigma_{\varpi}$ & $d$ & $d_{+}$ & $d_{-}$ & RC\\
  & [deg] & [deg] & [deg] &   & [mas\,yr$^{-1}$] & [mas\,yr$^{-1}$] & [mas\,yr$^{-1}$] & [mas\,yr$^{-1}$] & [mas] & [mas] & [pc] & [pc] & [pc] & \\
	\hline
 \multicolumn{15}{c}{...} \\

Gulliver~1 & 161.582 & -57.034 & 0.089 & 107 & -7.926 & 0.076 & 3.582 & 0.081 & 0.323 & 0.037 & 2837.3 & 2210.6 & 3963.2    & Y \\
Gulliver~2 & 122.883 & -37.404 & 0.073 & 67 & -4.951 & 0.095 & 4.576 & 0.119 & 0.696 & 0.056 & 1379.2 & 1212.1 & 1600.4    & N \\
Gulliver~3 & 122.536 & -37.244 & 0.035 & 47 & -2.962 & 0.086 & 4.106 & 0.109 & 0.191 & 0.073 & 4550.30 & 3127.0 & 8345.1    & N \\
Gulliver~4 & 122.164 & -37.5 & 0.079 & 64 & -2.912 & 0.061 & 3.033 & 0.062 & 0.30 & 0.04 & 3042.1 & 2332.3 & 4372.9    & Y \\
Gulliver~5 & 132.626 & -45.509 & 0.107 & 27 & -5.102 & 0.034 & 4.904 & 0.071 & 0.406 & 0.026 & 2297.8 & 1868.4 & 2981.3    & N \\
Gulliver~6 & 83.278 & -1.652 & 0.517 & 343 & -0.007 & 0.39 & -0.207 & 0.365 & 2.367 & 0.109 & 417.3 & 400.6 & 435.5    & N \\
Gulliver~7 & 141.746 & -55.127 & 0.034 & 90 & -3.547 & 0.151 & 3.108 & 0.100 & 0.084 & 0.048 & 8844.8 & 4693.3 & $\infty$    & N \\
Gulliver~8 & 80.56 & 33.792 & 0.102 & 38 & -0.156 & 0.225 & -2.982 & 0.152 & 0.872 & 0.087 & 1110.30 & 999.3 & 1249.0    & Y \\
Gulliver~9 & 126.998 & -47.929 & 0.969 & 265 & -5.992 & 0.272 & 6.915 & 0.375 & 1.985 & 0.092 & 496.5 & 473.0 & 522.4    & N \\
Gulliver~10 & 123.09 & -38.676 & 0.183 & 44 & -4.443 & 0.24 & 4.965 & 0.154 & 1.652 & 0.083 & 594.9 & 561.5 & 632.6    & N \\
Gulliver~11 & 67.996 & 43.62 & 0.131 & 64 & 0.400 & 0.206 & -2.229 & 0.149 & 1.061 & 0.068 & 917.3 & 840.2 & 1009.9    & N \\
Gulliver~12 & 181.174 & -61.308 & 0.076 & 52 & -5.948 & 0.071 & -0.41 & 0.088 & 0.559 & 0.034 & 1699.4 & 1452.6 & 2047.4    & N \\
Gulliver~13 & 104.858 & -13.254 & 0.115 & 78 & -2.941 & 0.125 & 0.281 & 0.113 & 0.62 & 0.061 & 1540.4 & 1334.9 & 1820.8    & Y \\
Gulliver~14 & 259.928 & -36.785 & 0.184 & 39 & -3.719 & 0.073 & -4.792 & 0.085 & 0.745 & 0.038 & 1291.5 & 1143.8 & 1483.1    & Y \\
Gulliver~15 & 272.599 & -16.723 & 0.089 & 84 & -1.06 & 0.119 & -1.638 & 0.100 & 0.506 & 0.067 & 1869.3 & 1574.8 & 2299.2    & N \\
Gulliver~16 & 23.433 & 60.751 & 0.046 & 57 & -1.249 & 0.058 & -0.614 & 0.083 & 0.208 & 0.065 & 4217.7 & 2964.6 & 7288.6    & Y \\
Gulliver~17 & 302.654 & 35.871 & 0.063 & 116 & -1.078 & 0.121 & -3.029 & 0.147 & 0.555 & 0.044 & 1711.8 & 1461.7 & 2065.0    & N \\
Gulliver~18 & 302.905 & 26.532 & 0.124 & 204 & -3.198 & 0.089 & -5.646 & 0.100 & 0.613 & 0.055 & 1558.6 & 1348.4 & 1846.3    & N \\
Gulliver~19 & 344.19 & 61.106 & 0.157 & 145 & 0.893 & 0.128 & -2.258 & 0.148 & 0.634 & 0.059 & 1507.9 & 1310.3 & 1775.7    & N \\
Gulliver~20 & 273.736 & 11.082 & 0.704 & 55 & 1.039 & 0.251 & -6.525 & 0.169 & 2.347 & 0.078 & 420.9 & 403.9 & 439.4    & N \\
Gulliver~21 & 106.961 & -25.462 & 0.364 & 126 & -1.929 & 0.118 & 4.205 & 0.141 & 1.504 & 0.05 & 652.2 & 612.3 & 697.8    & N \\
Gulliver~22 & 84.848 & 26.368 & 0.119 & 27 & -1.523 & 0.294 & -4.605 & 0.12 & 1.257 & 0.112 & 777.8 & 721.7 & 843.4    & N \\
Gulliver~23 & 304.255 & 38.055 & 0.046 & 150 & -2.446 & 0.087 & -4.444 & 0.095 & 0.246 & 0.05 & 3643.0 & 2670.1 & 5723.3    & Y \\
Gulliver~24 & 1.161 & 62.835 & 0.101 & 86 & -3.241 & 0.096 & -1.57 & 0.088 & 0.636 & 0.05 & 1504.9 & 1308.1 & 1771.6    & N \\
Gulliver~25 & 52.011 & 45.152 & 0.331 & 45 & 0.96 & 0.108 & -4.089 & 0.103 & 0.711 & 0.049 & 1351.3 & 1190.4 & 1562.4    & N \\
Gulliver~26 & 80.689 & 35.27 & 0.076 & 65 & 2.018 & 0.166 & -2.874 & 0.131 & 0.36 & 0.079 & 2570.1 & 2044.5 & 3459.1    & Y \\
Gulliver~27 & 146.088 & -54.117 & 0.049 & 65 & -4.658 & 0.066 & 3.467 & 0.079 & 0.322 & 0.027 & 2850.4 & 2218.1 & 3987.1    & N \\
Gulliver~28 & 293.559 & 18.059 & 0.555 & 68 & -4.485 & 0.158 & -3.4 & 0.136 & 1.581 & 0.065 & 621.3 & 584.9 & 662.5    & N \\
Gulliver~29 & 256.745 & -35.205 & 0.679 & 440 & 1.325 & 0.161 & -2.206 & 0.174 & 0.905 & 0.061 & 1070.9 & 967.3 & 1199.3    & N \\
Gulliver~30 & 313.673 & 45.996 & 0.083 & 63 & -2.524 & 0.07 & -3.697 & 0.066 & 0.427 & 0.043 & 2192.3 & 1797.7 & 2808.6    & N \\
Gulliver~31 & 301.912 & 38.232 & 0.077 & 54 & -1.50 & 0.063 & -3.113 & 0.086 & 0.396 & 0.033 & 2352.8 & 1905.6 & 3077.1    & N \\
Gulliver~32 & 98.383 & 7.478 & 0.10 & 39 & -0.879 & 0.104 & 2.304 & 0.107 & 0.577 & 0.061 & 1649.8 & 1416.2 & 1975.6    & N \\
Gulliver~33 & 318.159 & 46.345 & 0.31 & 98 & 0.274 & 0.12 & -3.959 & 0.123 & 0.867 & 0.063 & 1116.5 & 1004.6 & 1256.9    & N \\
Gulliver~34 & 167.722 & -59.158 & 0.045 & 31 & -6.179 & 0.059 & 2.409 & 0.072 & 0.223 & 0.034 & 3972.3 & 2842.1 & 6580.6    & N \\
Gulliver~35 & 150.379 & -58.198 & 0.09 & 34 & -6.973 & 0.111 & 3.967 & 0.092 & 0.405 & 0.025 & 2301.9 & 1871.4 & 2991.6 & N \\
Gulliver~36 & 123.185 & -35.111 & 0.14 & 101 & -0.236 & 0.078 & 0.609 & 0.071 & 0.725 & 0.042 & 1326.7 & 1171.3 & 1529.6    & Y \\
Gulliver~37 & 292.077 & 25.347 & 0.105 & 59 & -0.775 & 0.074 & -3.74 & 0.092 & 0.642 & 0.038 & 1490.9 & 1297.6 & 1751.8    & Y \\
Gulliver~38 & 300.808 & 34.435 & 0.06 & 111 & -0.921 & 0.117 & -2.594 & 0.134 & 0.4 & 0.043 & 2329.1 & 1889.9 & 3036.6    & Y \\
Gulliver~39 & 163.697 & -58.05 & 0.045 & 42 & -4.407 & 0.07 & 1.274 & 0.06 & 0.335 & 0.035 & 2747.5 & 2155.3 & 3788.6    & Y \\
Gulliver~40 & 163.095 & -58.394 & 0.082 & 36 & -7.686 & 0.044 & 2.476 & 0.061 & 0.589 & 0.041 & 1618.7 & 1393.1 & 1931.3    & N \\
Gulliver~41 & 277.718 & -12.429 & 0.032 & 59 & -1.79 & 0.297 & -4.709 & 0.257 & 0.171 & 0.132 & 5000.2 & 3333.8 & 9982.9    & N \\
Gulliver~42 & 303.935 & 37.851 & 0.054 & 83 & -2.778 & 0.189 & -5.452 & 0.217 & 0.181 & 0.074 & 4763.9 & 3226.7 & 9092.1    & N \\
Gulliver~43 & 296.283 & 24.558 & 0.075 & 79 & -2.922 & 0.085 & -5.796 & 0.104 & 0.351 & 0.051 & 2631.6 & 2083.3 & 3571.7    & N \\
Gulliver~44 & 127.249 & -38.095 & 0.189 & 153 & -0.666 & 0.136 & 2.301 & 0.132 & 0.785 & 0.053 & 1228.2 & 1093.9 & 1400.2    & Y \\
Gulliver~45 & 104.617 & 3.104 & 0.054 & 50 & -0.76 & 0.201 & 2.51 & 0.19 & 0.284 & 0.085 & 3196.9 & 2422.4 & 4699.2    & N \\
Gulliver~46 & 186.234 & -61.973 & 0.024 & 80 & -6.946 & 0.093 & 0.125 & 0.07 & 0.181 & 0.055 & 4766.3 & 3227.8 & 9113.6    & N \\
Gulliver~47 & 107.932 & 0.827 & 0.161 & 87 & -1.957 & 0.093 & -0.432 & 0.076 & 0.491 & 0.05 & 1922.2 & 1612.2 & 2378.2  & N \\
Gulliver~48 & 316.334 & 50.733 & 0.28 & 104 & -4.781 & 0.141 & -6.671 & 0.166 & 1.058 & 0.054 & 919.8 & 842.3 & 1013.0    & N \\
Gulliver~49 & 350.704 & 61.988 & 0.156 & 165 & -4.022 & 0.112 & -3.05 & 0.105 & 0.587 & 0.039 & 1621.7 & 1395.5 & 1936.5    & N \\
Gulliver~50 & 181.362 & -62.678 & 0.106 & 76 & -7.204 & 0.102 & 1.663 & 0.063 & 0.514 & 0.042 & 1841.7 & 1554.9 & 2256.8    & N \\
Gulliver~51 & 30.335 & 63.801 & 0.075 & 41 & -4.892 & 0.097 & -0.149 & 0.079 & 0.647 & 0.026 & 1479.6 & 1288.9 & 1736.5    & Y \\
Gulliver~52 & 161.669 & -59.508 & 0.153 & 62 & -4.755 & 0.068 & 1.47 & 0.08 & 0.396 & 0.042 & 2350.8 & 1903.3 & 3073.5    & N \\
Gulliver~53 & 80.975 & 34.012 & 0.123 & 36 & 0.401 & 0.109 & -2.837 & 0.106 & 0.384 & 0.038 & 2421.9 & 1949.5 & 3196.2    & Y \\
Gulliver~54 & 81.297 & 33.688 & 0.102 & 35 & -0.595 & 0.140 & -7.299 & 0.175 & 0.791 & 0.061 & 1219.0 & 1086.5 & 1388.3    & N \\
Gulliver~55 & 297.833 & 18.661 & 0.085 & 34 & -2.043 & 0.08 & -2.53 & 0.072 & 0.369 & 0.026 & 2513.2 & 2008.3 & 3354.3      & Y \\
Gulliver~56 & 95.38 & 26.909 & 0.071 & 26 & 0.534 & 0.078 & -3.24 & 0.071 & 0.458 & 0.04 & 2054.9 & 1704.5 & 2586.6    & N   \\
Gulliver~57 & 141.203 & -48.075 & 0.11 & 106 & -6.737 & 0.154 & 4.085 & 0.123 & 0.701 & 0.056 & 1370.5 & 1205.3 & 1588.1    & N   \\
Gulliver~58 & 191.515 & -61.965 & 0.074 & 205 & -3.592 & 0.125 & -0.353 & 0.116 & 0.398 & 0.055 & 2344.2 & 1898.9 & 3062.3      & Y \\
Gulliver~59 & 195.721 & -64.6 & 0.172 & 78 & -2.408 & 0.101 & -1.053 & 0.087 & 0.434 & 0.039 & 2161.7 & 1777.5 & 2758.0    & N   \\
Gulliver~60 & 303.436 & 29.672 & 0.109 & 129 & 1.752 & 0.092 & 0.219 & 0.117 & 0.899 & 0.037 & 1077.1 & 972.4 & 1207.2    & N   \\

 \multicolumn{15}{c}{...} \\
	\hline
	\hline
	\end{tabular}
\tablefoot{$N$: number of stars with membership probabilities over 50\%. $d$: mode of the distance likelihood. $d_{+}$ and $d_{-}$: modes obtained when adding (respectively subtracting) 0.1\,mas to (from) all parallaxes. RC: indicates whether (Y) or not (N) the CMD features red clump stars with membership probabilities greater than 50\%. }
\end{center}
\end{table*}

We also report in Table~\ref{table:meanparams} the radius r$_{50}$ (in degrees) containing 50\% of the cluster members. We show this parameter as a function of the mean cluster parallax in Fig.~\ref{fig:r50_vs_par}. This parameter is not meant to be a physically accurate description of the cluster extension, as the field of view employed for every individual cluster may or may not contain its most external region, and should be taken as an indication of the area in which cluster members are detectable with our method. Experimenting with a sparse cluster (ESO~130~06) and a massive cluster (NGC~6705) of comparable apparent sizes (r$_{50}$=0.109$^{\circ}$ and 0.074$^{\circ}$, respectively) we found in both cases that using fields of view of radius 0.2 to 0.6 degrees could change the value of r$_{50}$ by up to $20\%$, while the median proper motions and parallaxes varied by less than 0.02\,mas\,yr$^{-1}$ and 0.01\,mas. 

It is also well-known that clusters exhibiting mass segregation have significantly different sizes depending on the magnitude of the stars considered \citep[see e.g.][]{Allison09,CantatGaudin14m11,Dib18}. Although the most common approach to estimating the size of an OC is through the fitting of a density profile,  other methods have been suggested, such as establishing the radius that provides the best contrast between field and cluster stars in astrometric space \citep{Sanchez18}. 
A better estimate of the true extent of a cluster (and identification of its most distant members) could be obtained by modeling the background distribution of the field stars, and considering the individual kinematics of each star, as for instance done by \cite{Reino18} for the Hyades clusters.

\begin{figure}[ht]
\begin{center} \resizebox{\hsize}{!}{\includegraphics[scale=0.5]{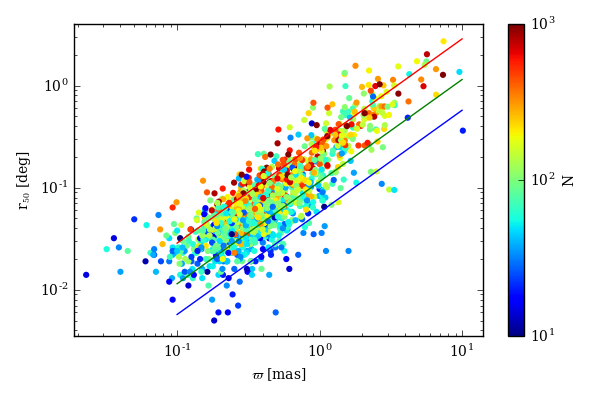}} \caption{\label{fig:r50_vs_par} Radius r$_{50}$ containing 50\% of the cluster members identified in this study, against mean parallax of the cluster. The colour code indicates the number of members. The blue, green, and red solid lines indicate the angular sizes corresponding to physical radii of 1\,pc, 2\,pc, and 5\,pc (respectively). } \end{center}
\end{figure}

We also report in Table~\ref{table:meanparams} whether the CMD of the clusters present possible red clump stars with membership probabilities greater than 50\%. We recall that the membership assignment procedure applied in this study does not rely on any photometric criteria and does not take into account radial velocities. Those potential red clump stars might therefore not all be true members.

\subsection{Obtaining distances from parallaxes} \label{sec:distances}
We have estimated distances to the clusters through a maximum likelihood procedure, maximising the quantity:

\begin{equation}
	\mathcal{L} \propto \prod_{i=1} P(\varpi_i|d,\sigma_{\varpi_i})  = \prod_{i=1} \frac{1}{ \sqrt{2 \pi \sigma_{\varpi_i}^2} }  \exp \left( - \frac{  (\varpi_i - \frac{1}{d}) ^2  }{ 2 \sigma_{\varpi_i}^2 } \right)
\end{equation} 

\noindent where $P(\varpi_i|d,\sigma_{\varpi_i})$ (which is Gaussian and symmetrical in $\varpi$ but not in $d$) is the probability of measuring a value of $\varpi_i$ (in mas) for the parallax of star $i$, if its true distance is $d$ (in kpc) and its measurement uncertainty is $\sigma_{\varpi_i}$. 
We here neglect correlations between parallax measurements of all stars, and consider the likelihood for the cluster distance to be the product of the individual likelihoods of all its members. This approach also neglects the intrinsic physical depth of a cluster by assuming all its members are at the same distance. This approximation holds true for the distant clusters, whose depth (expressed in mas) is much smaller than the individual parallax uncertainties, but might not be optimal for the most nearby clusters.

As reported in \citet{Lindegren18} and \citet{Arenou18} \citep[and confirmed by][]{Riess18,Zinn18,Stassun18}, the \textit{Gaia}~DR2 parallaxes are affected by a zero-point offset. Following the guidelines of \citet{Lindegren18} and \citet{Luri18}, we accounted for this bias by adding +0.029\,mas to all parallaxes before performing our distance estimation.

We report in Table~\ref{table:meanparams} the mode of the likelihood, as well as the distances $d_{16}$ and $d_{84}$ defining the 68\% confidence interval, and $d_{5}$ and $d_{95}$ defining the 90\% confidence interval\footnote{The values of $d_{5}$, $d_{16}$, $d_{84}$, and $d_{95}$ are only reported in the electronic table}. In addition to the global zero point already mentioned, local systematics possibly reaching 0.1\,mas are still present in \textit{Gaia}~DR2 parallaxes \citep{Lindegren18}. To provide a bracketing of the possible distances of the most infortunate cases, we also provide the modes $d_+$ and $d_-$ of the likelihoods obtained adding $\pm0.1$\,mas to the parallaxes. 

For the large majority of objects in this study, the assumption that the stars are physically located at the same distance (and therefore have the same true parallax) leads to a small fractional uncertainty $\sigma_{\langle\varpi\rangle} / \langle\varpi\rangle$ on the mean parallax. Considering that the statistical uncertainty on the mean parallax of a cluster decreases with the square root of the number of stars, 84\% of the OCs in our study have fractional errors below 5\% (94\% have fractional uncertainties below 10\%). For those clusters, inverting the mean parallax provides a reasonable estimate of the distance. The presence of a systematic bias however makes the accuracy of the mean parallax much poorer than the statistical precision, and the range of possible distances is better estimated through a maximum likelihood approach. For the most distant clusters, the distance estimate when subtracting 0.1\,mas to the parallaxes diverges to infinity. In the presence of this unknown local bias, the distances to clusters with mean parallaxes smaller than $\sim0.2$\,mas would be better constrained by a Bayesian approach using priors based on an assumed density distribution of the Milky Way \citep[as in e.g.][]{BailerJones18} or photometric considerations \citep[e.g.][]{Anderson17}, or simply with more classical isochrone fitting methods.

\subsection{Comparison with the literature}
We compared the astrometric parameters obtained in this study with those given in \citet{GDR2hrd} for the 38 OCs in common. We find an excellent agreement between the two sets of results, with a typical difference in mean parallax under 0.02\,mas, and under 0.05\,mas\,yr$^{1}$ in proper motions (see Fig.~\ref{fig:compareAstro}). The largest differences correspond to the most nearby clusters with the largest apparent dispersions in astrometric parameters.

\begin{figure}[ht]
\begin{center} \resizebox{\hsize}{!}{\includegraphics[scale=0.5]{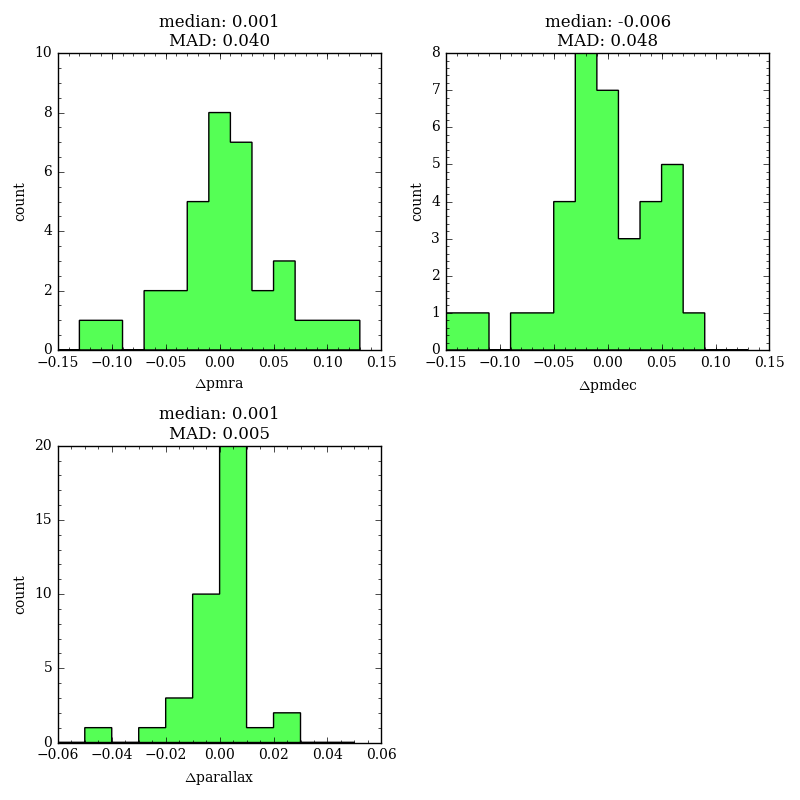}} \caption{\label{fig:compareAstro} Value of the mean astrometric parameters minus the value quoted in \citet{GDR2hrd} for the 38 OCs in common between both studies. } \end{center}
\end{figure}

We also made comparisons with the distances to the clusters of the BOCCE project \citep{Bragaglia06}. Figure~\ref{fig:bocce_expected_par} shows the difference between our parallax determination (uncorrected) and the expected parallax given the literature distance of the cluster. We remark a significant median zero point of -0.048\,mas (-48\,$\mu$as), slightly more negative than the -0.029\,mas value we adopted from \citep{Lindegren18}. This value is compatible with the independent determinations of \citet{Riess18}, \citet{Stassun18}, and \citet{Zinn18}, who determined zero-points of -46\,$\mu$as, -82\,$\mu$as, and -50\,$\mu$as respectively. It is also compatible with the values quoted by \citet{Arenou18} who assessed the zero point with a variety of reference tracers. We however refrain from drawing strong conclusions as to the value of the zero point from the small number of BOCCE clusters, and for the rest of this study we only corrected the parallaxes for the 29\,$\mu$as negative zero point mentioned in \citet{Lindegren18}.

\begin{figure}[ht]
\begin{center} \resizebox{\hsize}{!}{\includegraphics[scale=0.5]{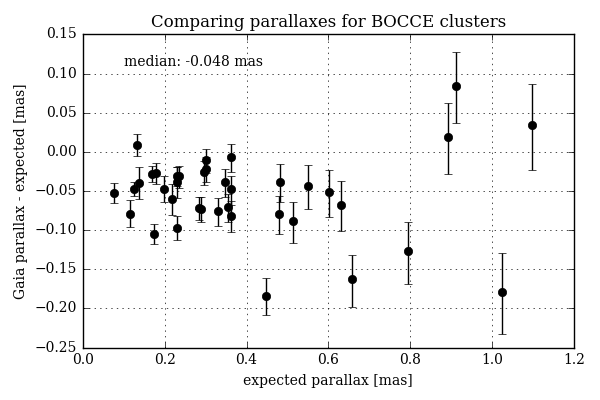}} \caption{\label{fig:bocce_expected_par} Difference between our parallax determination and the expected value given their distance, for the BOCCE clusters. The error bars represent the quadratic sum of our uncertainty and a 5\% error on the reference distance. References are listed in Table~\ref{table:bocce}. } \end{center}
\end{figure}

\begin{table}
\begin{center}
	\caption{ \label{table:bocce} Distances to the BOCCE clusters.}
	\small\addtolength{\tabcolsep}{-1pt}
	\begin{tabular}{c c c | c c c}
	\hline
	\hline
   	& \multicolumn{2}{c}{This study} & \multicolumn{3}{c}{BOCCE}        \\
   OC  	& $\varpi$      &  $d$         & $d$      &     $\varpi$    & Ref.  \\
    	& [mas]         & [pc]           & [pc]   &   [mas]               \\
	\hline

Berkeley~17 & 0.281 & 3180 & 2754 & 0.363 & B06 \\	
Berkeley~20 & 0.036 & 12232 & 8710 & 0.115 & A11 \\	
Berkeley~21 & 0.152 & 5211 & 5012 & 0.200 & BT06 \\	
Berkeley~22 & 0.069 & 9065 & 5754 & 0.174 & DF05 \\	
Berkeley~23 & 0.151 & 5365 & 5623 & 0.178 & C11 \\	
Berkeley~27 & 0.190 & 4339 & 4365 & 0.229 & D12 \\	
Berkeley~29 & 0.023 & 15137 & 13183 & 0.076 & BT06 \\	
Berkeley~31 & 0.141 & 5655 & 7586 & 0.132 & C11  \\	
Berkeley~32 & 0.280 & 3202 & 3311 & 0.302 & T07 \\	
Berkeley~34 & 0.098 & 7016 & 7244 & 0.138 & D12 \\	
Berkeley~36 & 0.203 & 4220 & 4266 & 0.234 & D12  \\	
Berkeley~66 & 0.158 & 5029 & 4571 & 0.219 & A11  \\	
Berkeley~81 & 0.255 & 3454 & 3020 & 0.331 & D14 \\	
Collinder~110 & 0.424 & 2201 & 1950 & 0.513 & BT06 \\	
Collinder~261 & 0.315 & 2894 & 2754 & 0.363 & BT06  \\	
King~11 & 0.263 & 3386 & 2239 & 0.447 & T07 \\	
King~8 & 0.132 & 5937 & 4365 & 0.229 & C11      \\	
NGC~1817 & 0.551 & 1718 & 1660 & 0.602 & D14 \\	
NGC~2099 & 0.667 & 1434 & 1259 & 0.794 & BT06  \\	
NGC~2141 & 0.198 & 4359 & 4365 & 0.229 & D14 \\	
NGC~2168 & 1.131 & 861 & 912 & 1.096 & BT06 \\	
NGC~2243 & 0.211 & 4143 & 3532 & 0.283 & BT06  \\	
NGC~2323 & 0.997 & 973 & 1096 & 0.912 & BT06 \\	
NGC~2355 & 0.495 & 1897 & 1520 & 0.658 & D15	      \\	
NGC~2506 & 0.291 & 3112 & 3311 & 0.302 & BT06 \\	
NGC~2660 & 0.308 & 2953 & 2884 & 0.347 & BT06 \\	
NGC~2849 & 0.142 & 5724 & 5888 & 0.170 & A13 \\	
NGC~3960 & 0.399 & 2326 & 2089 & 0.479 & BT06 \\	
NGC~6067 & 0.442 & 2116 & 2080 & 0.481 & --  	      \\	
NGC~6134 & 0.845 & 1142 & 977 & 1.024 & A13 \\	
NGC~6253 & 0.563 & 1683 & 1585 & 0.631 & BT06  \\	
NGC~6709 & 0.912 & 1060 & 1120 & 0.893 & --  	      \\	
NGC~6819 & 0.356 & 2595 & 2754 & 0.363 & BT06 \\	
NGC~6939 & 0.506 & 1864 & 1820 & 0.549 & BT06 \\	
NGC~7790 & 0.269 & 3333 & 3388 & 0.295 & BT06 \\	
Pismis~2 & 0.215 & 4011 & 3467 & 0.288 & BT06  \\	
Tombaugh~2 & 0.079 & 8945 & 7943 & 0.126 & A11  \\	
Trumpler~5 & 0.284 & 3185 & 2820 & 0.355 & D15 \\	

	\hline
	\hline
	\end{tabular}
\tablefoot{References: B06: \citet{Bragaglia06be17}; A11: \citet{Andreuzzi11}; BT06: \citet{Bragaglia06}; DF05: \cite{DiFabrizio05}; C11: \citet{Cignoni11}; D12: \citet{Donati12}; T07: \citet{Tosi07}; D14: \citet{Donati14bocce}; D15: \citet{Donati15}; A13: \citet{Ahumada13}. For NGC~6067 and NGC~6709: private communication with P. Donati.}
\end{center}
\end{table}

\begin{table}
\begin{center}
	\caption{ \label{table:NP} Number of clusters with at least $N$ stars having membership probabilities $\geq p$.}
	\small\addtolength{\tabcolsep}{-0pt}
	\begin{tabular}{ c | c  c  c  c  c  c  c }
	\hline
	\hline

 & \multicolumn{6}{c}{$N$} \\
$p$ & 5 & 10 & 20 & 50 & 100 & 200 & 500 \\
	\hline

10 & 1229 & 1229 & 1226 & 1176 & 991 & 646 & 208 \\
30 & 1229 & 1229 & 1224 & 1126 & 867 & 447 & 130 \\
50 & 1229 & 1228 & 1216 & 1029 & 681 & 324 & 98 \\
70 & 1222 & 1205 & 1137 & 818 & 501 & 232 & 70 \\
90 & 1119 & 1051 & 903 & 564 & 323 & 157 & 44 \\

	\hline
	\hline
	\end{tabular}

\end{center}
\end{table}

We also attempted a comparison of the distances derived in this study with those listed in the MWSC catalogue. We generally find a good agreement for the clusters with more than a few hundred members, but for most clusters with fewer members the discrepancies can be significant. We remark that the literature values themselves can vary a lot between sources (for instance Berkeley~76 is listed at 2360\,pc in MWSC and 12600\,pc in DAML), and our distance estimate might be in agreement with both, one, or none of these values, with discrepancies too large to be explained by instrumental errors (e.g. NGC~2509, Fig.~\ref{fig:distance_discrepancy}). 
We also remark that on average, our distances estimated from parallaxes tend to agree more often with the more distant literature value. 
We suggest that the isochrone fitting procedure from which most photometric distances are estimated might be biased in many of those instances, as they might be affected by field contamination, as well as the degeneracies between distance, reddening, and metallicity. The strength of the \textit{Gaia} data does not only reside in parallaxes from which distances can be inferred, and in precise and deep photometry, but also in our ability to obtain clean colour-magnitude diagrams from which astrophysical parameters can be infered.

\begin{figure}[ht]
\begin{center} \resizebox{\hsize}{!}{\includegraphics[scale=0.5]{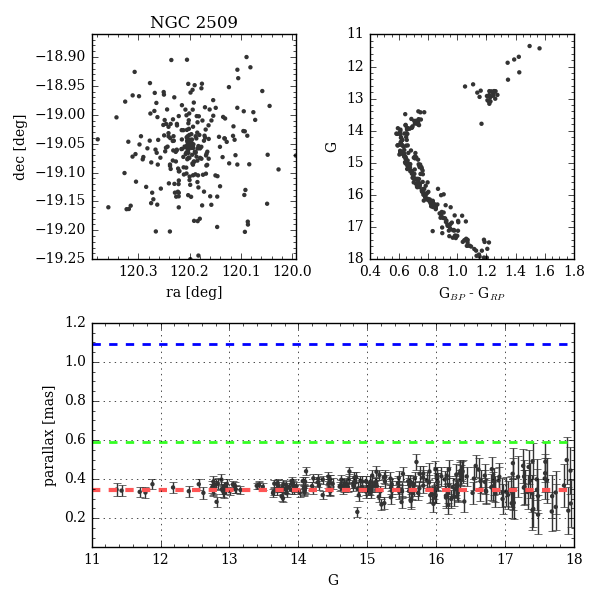}} \caption{\label{fig:distance_discrepancy} Sky distribution (top left), colour-magnitude diagram (top right), and parallax as a function of magnitude (bottom) for the members of NGC~2509. The median parallax is 0.36\,mas. Available literature distances for this cluster are 912\,pc (DAML, blue), 1700pc (MWSC, green) and 2900\,pc \citep[][red]{CarraroCosta07}, which correspond to parallaxes of 1.09\,mas, 0.59\,mas, and 0.35\,mas (respectively).} \end{center}
\end{figure}

\section{Specific remarks on some clusters} \label{sec:remarks}
It would be lengthy and impractical to devote a section to every object under analysis in this study, but we provide additional comments on the specific cases of two clusters classified as OCs that are likely globular clusters, and on the previously unreported OCs discovered in this study.

\subsection{BH~140 and FSR~1758 are globular clusters}
The colour-magnitude diagrams of BH~140 and FSR~1758 (shown in Fig.~\ref{fig:BH_140_cmd}) present the typical aspect of globular clusters, with a prominent giant branch and an interrupted horizontal branch displaying a gap in the locus occupied by RR Lyrae. They are also clearly seen as rich and regular distributions on the sky. These two objects, located near the Galactic plane ($b$=-4.3$^{\circ}$ and -3.3$^{\circ}$ for BH~140 and FSR~1758, respectively) also exhibit small parallaxes ($\varpi$=0.16\,mas and 0.09\,mas), so their distances cannot be estimated accurately from parallaxes alone. FSR~1758, with a Galactic longitude $l=349.2^{\circ}$, seems to be located deep in the inner disk, as the distance estimated from its parallax yields a Galactocentric radius $R_{\mathrm{GC}}=1560$\,pc, and altitude Z=-470\,pc.

The broad appearance of the CMD of BH~140 can be explained by blended photometry in the inner regions. As discussed in \citet{Evans18} and illustrated in \citet{Arenou18}, the $G_{BP}$ and $G_{RP}$ fluxes of \textit{Gaia}~DR2 sources might be overestimated as a result of background contamination, and the effect is especially relevant in crowded fields such as the core of globular clusters.

The bottom panels of Fig.~\ref{fig:BH_140_cmd} show that their distinct proper motions allow us to separate the cluster members from the field stars, picking 434 probable members out of 13\,000 sources in the case of BH~140, and 540 probable members out of more than 120\,000 sources in the very populated field of FSR~1758. We verified that both objects are absent from the updated web page of the globular cluster catalogue of \citet{Harris96}.

\begin{figure}[ht]
\begin{center} \resizebox{\hsize}{!}{\includegraphics[scale=0.5]{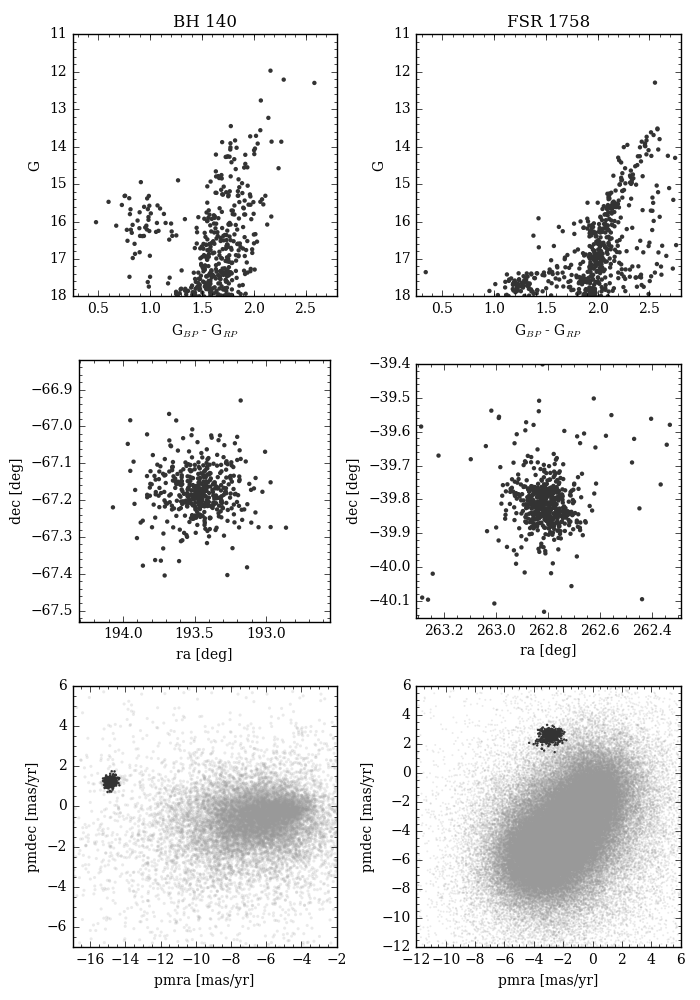}} \caption{\label{fig:BH_140_cmd} Top left: colour-magnitude diagram for BH~140. Middle left: distribution on the sky for the same stars. Bottom left: proper motion diagram showing the field in light grey, and the cluster stars as black dots. Right column: same for FSR~1758.} \end{center}
\end{figure}

\subsection{Newly found clusters} \label{sec:gulliver}

We report on the serendipitous discovery of 60 hitherto unreported candidate clusters.
These clusters, hereafter named ``Gulliver'', were found investigating the known OCs. They were identified as groups of stars with consistent proper motions and parallaxes, and distributions on the sky significantly more concentrated than a uniform distribution. We manually verified that their CMDs present aspects compatible with them being single stellar populations, but did not perform any additional comparisons with stellar isochrones. 

All of them were found in the same field of view as a known OC under investigation, but have distinct proper motions and parallaxes and a distinct aspect in a colour-magnitude diagram, therefore are not necessarily related. The location, proper motions, parallaxes and colour-magnitude diagram of Gulliver~1 are shown in Fig.~\ref{fig:gulliver1_4panels}, as an example. 

\begin{figure}[ht]
\begin{center} \resizebox{\hsize}{!}{\includegraphics[scale=0.5]{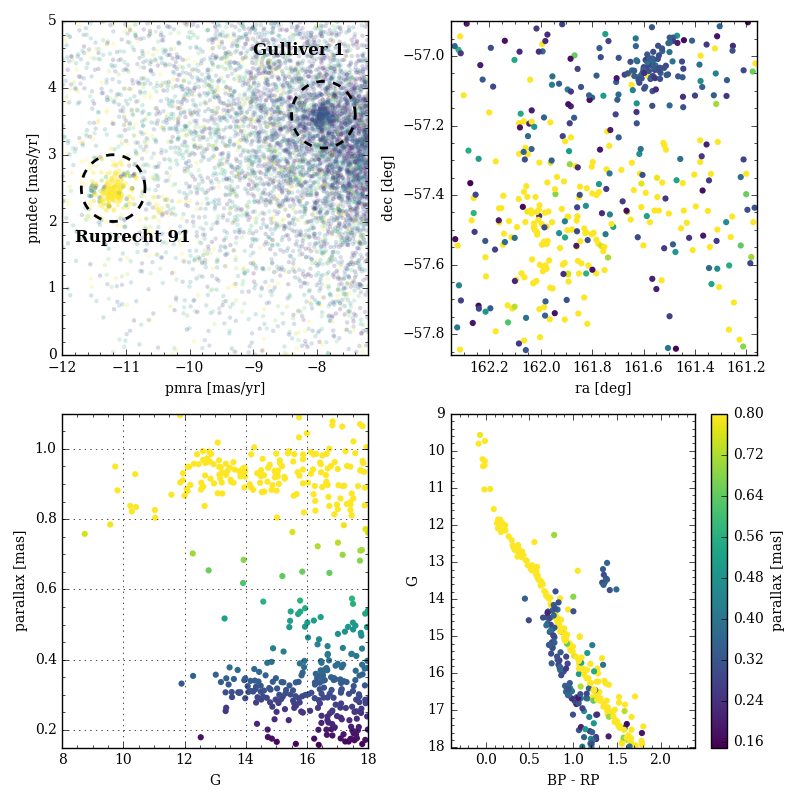}} \caption{\label{fig:gulliver1_4panels} Top left: Ruprecht~91 and Gulliver~1 seen as overdensities in a proper motion diagram. Top right: location of the stars selected from their proper motions. Bottom left: parallax against $G$ magnitude for those selected stars. Bottom right: CMD of those selected stars. The selection used in this figure was performed on proper motions only for illustration purposes. } \end{center}
\end{figure}

The estimated distances of those proposed new clusters range from 415 to 8800\,pc, and 32 of them are within 2\,kpc of the Sun. They are not located in a specific region of the sky, but rather seem randomly distributed along the Galactic plane. Their coordinates and parameters are listed in Table~\ref{table:meanparams}, and their distribution projected on the Galactic plane in Fig.~\ref{fig:map_XY_spiral_ages}.

\subsection{The non-detected clusters} \label{sec:notfound}
The total number of objects we were able to clearly identify is less than 50\% of the initial list of clusters and candidates. Combinations of many factors can cause a cluster to be difficult to detect, such as the source density of the background, interstellar extinction, how populated a cluster is, its age, or how its proper motions differ from the field stars. The literature also lists objects flagged as dubious clusters or as asterisms, albeit with some disagreements between DAML and MWSC\footnote{In this study we were able to detect Alessi~17, Pismis~21, King~25, and NGC~1724, four OCs flagged as probable asterisms in the DAML catalogue.}. We removed from the list of non-detected clusters those flagged as \textit{asterism}, \textit{dubious}, \textit{infrared}, or \textit{embedded}, and avoided classifying as non-detection those found under a different name\footnote{We used the list of multiple names provided at \url{www.univie.ac.at/webda/double_names.html}, to which we added, based on the proximity in location and proposed distance: Teutsch~1 = Koposov~27; Juchert~10 = FSR~1686; Roslund~6 = RSG~6; DBSB~5 = Mayer~3; PTB~9 = NGC~7762; ESO~275~01 = FSR~1723; S1 = Berkeley~6; Pismis~24 = NGC~6357.}.

Figure~\ref{fig:ebv_vs_dist_histograms} shows the distribution of distances and extinction $E\mathrm{(B-V)}$ (as listed in MWSC) for the sample of OCs we detected and those we did not. Both appear very similar, except at the extreme values of distance and extinction. We only detect the OCs more distant than 10\,kpc if their extinction is lower than $E\mathrm{(B-V)}=1.5$. The only distant low-extinction cluster we do not detect is the very old object Saurer~1 \citep[$\sim5$\,Gyr][]{Carraro04}, whose red giant stars are fainter than our chosen magnitude limit $G=18$. As an experiment, we manually selected the red giant stars of FSR~0190 \citep[][an object we failed to detect in this study]{Froebrich08} from their 2MASS photometry, and found that they all have $G$ magnitudes fainter than our limit. Due to their large proper motion and parallax uncertainties these stars do not stand as strong overdensities in astrometric space, but are clearly seen as a concentration on the sky, showing that a mixed approach taking into account photometry (in particular in infrared filters) would enable us to identify (and discover) more of those distant reddened objects than astrometry alone.

\begin{figure}[ht]
\begin{center} \resizebox{\hsize}{!}{\includegraphics[scale=0.5]{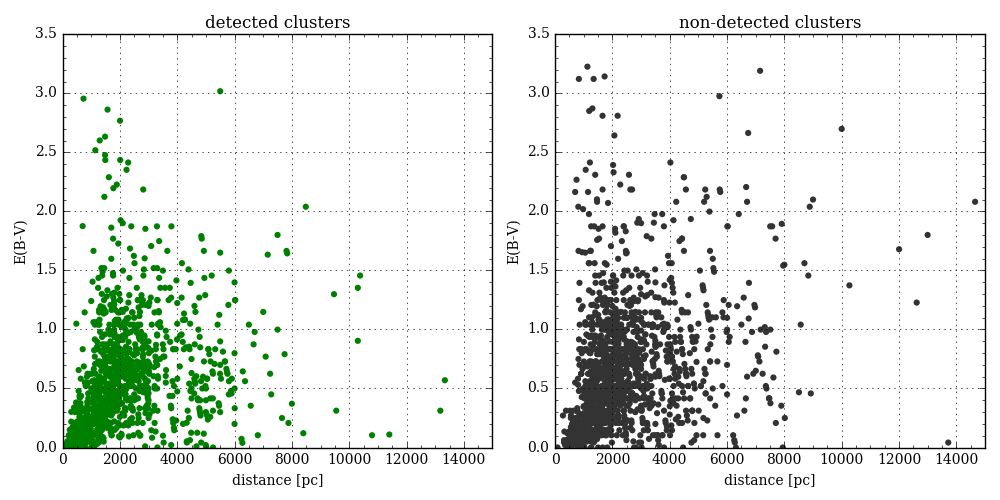}} \caption{\label{fig:ebv_vs_dist_histograms} Distance and extinction (as listed in MWSC) for the clusters we detected in this study (left) and those we did not (right). The second panel does not include the objects flagged as dubious or as asterisms (see text).} \end{center}
\end{figure}

  We remark large differences in detection rates between the various ``families'' of clusters: we detected over 90\% of the Alessi, Berkeley, IC, Pismis, and Trumpler OCs, 80\% of the NGC and BH OCs, $\sim$65\% of the BDSB and Ruprecht OCs, but found under 40\% of the ASCC clusters, 6\% of the Loden, and none of the 203 putative clusters listed in \citet{Schmeja14} and \citet{Scholz15}.  Those 203 objects are located at high Galactic latitudes in regions of low background density (see top panel of Fig.~\ref{fig:statsofnondetected}), most of them at distances under 2\,kpc, originally discovered with PPMXL proper motions, and should therefore easily be identified with \textit{Gaia}~DR2 astrometry. Furthermore, their proposed ages of $\log t > 8.5$ at Galactocentric distances of 6\,kpc to 8\,kpc and altitudes Z$\sim$400\,pc to 800\,pc are at odds with all studies reporting an apparent absence of old and of high-altitude OCs in the inner disk (see discussion in Sect.~\ref{sec:galaxy}). We applied our method to those 203 objects after performing an additional cut retaining only stars with magnitudes $G<15$, which given their expected cluster parameters should allow to see their trace with great contrast in astrometric space, with no success. Finally, for the 139 candidates of \citet{Schmeja14} we directly cross-matched their lists of probable members with the \textit{Gaia}~DR2 data with no magnitude restriction, and found that although their stars are consistent groups in PPXML proper motions, they are not consistent in \textit{Gaia}~DR2 astrometry (which has on average nominal errors 100 to 200 times smaller than PPXML) or photometry except in one case: the group of stars reported as MWSC~5058 belong to the dwarf galaxy Leo~I ($d\sim250$\,kpc), with proper motions and parallaxes near zero, and $G$-magnitudes fainter than $\sim18.6$, and is therefore not a Galactic cluster either. 

 We could only identify 15\% of all FSR clusters (a proportion unchanged if we limit the sample to $E\mathrm{(B-V)}<1$ and distances under 2\,kpc), fewer than expected since the authors \citep{Froebrich07} estimate a false positive rate of 50\%. We were not able to detect any of their inner-disk high-altitude candidates. After discarding the cluster candidates from these three studies \citep{Froebrich07,Schmeja14,Scholz15} it appears that most of the clusters we failed to detect are located at low Galactic latitudes and towards the inner disk (see bottom-left panel of Fig.~\ref{fig:statsofnondetected}), which correspond to regions of higher density and extinction. Rather than just the source density along the line of sight, the contrast between cluster and field stars in astrometric space is the relevant variable. For instance, in the case of BH~140 (illustrated in Fig.~\ref{fig:BH_140_cmd}), the proper motions of the cluster are very different from the field stars, which allows it to be clearly seen despite being in a crowded and reddened field of view. The majority of missing clusters are located between 1 and 2\,kpc. The known distant clusters are usually easier to identify, as most of them are located towards the Galactic anticentre and at high Galactic latitudes. 

\begin{figure}[ht]
\begin{center} \resizebox{\hsize}{!}{\includegraphics[scale=0.5]{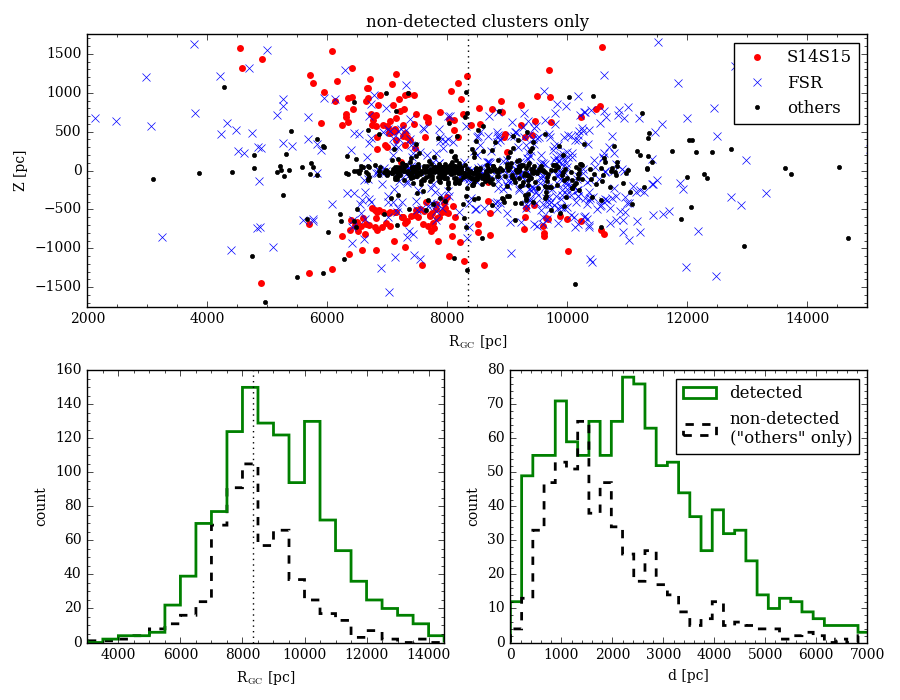}} \caption{\label{fig:statsofnondetected} Top panel: expected locations of the OCs we failed to detect in this study (excluding infrared clusters, dubious clusters and proposed asterisms). S14S15 indicates the candidates of \citet{Schmeja14} and \citet{Scholz15}, and FSR those of \citet{Froebrich07}. Bottom left: distribution in Galactocentric distance of the identified OCs (green) and those we failed to identify, excluding the candidates of S14S15 and FSR (dashed black). The vertical dotted line indicates the solar radius at 8.34\,kpc. Bottom right: same as previous panel, for the distribution in heliocentric distance.} \end{center}
\end{figure}

Given the unprecedented quality of the \textit{Gaia}~DR2 astrometry, it is likely that many of the objects we failed to detect are not true clusters. Concluding on the reality of those objects is however beyond the scope of this paper, as it requires to go through the individual lists of members (when their authors have made them public) or the data the discovery was based on. For instance, we do not find any trace of the cluster ASCC~35, an object for which \citet{Netopil15} already mention the absence of a visible sequence in a colour-magnitude diagram. We did not find Loden~1 either, for which the literature lists distances of 360 to 786\,pc and ages up to 2\,Gyr \cite[despite the original description by][of the group containing A-type main sequence stars]{Loden80}, and for which \citet{Han16} conclude that it is not a cluster based on measured radial velocities. Another historical example of an object listed as a cluster despite an absence of convincing observations is NGC~1746, originally listed in the New General Catalogue of Nebul{\ae} and Clusters of Stars \citep{Dreyer1888}. Its reality was questioned by \citet{Straizys92}, \citet{Galadi98i}, \citet{Tian98}, and \citet{Landolt10}, on the basis of photometry and astrometry, but NGC~1746 is still incuded in the MWSC catalogue. In this present study we found no trace of this cluster.

It is possible that other such objects discovered as coincidental groupings by manual inspection of photographic plates are not true stellar clusters, even those for which the literature lists ages or distances. Although it is more difficult to prove a negative (here the non-existence of a cluster) with a high degree of certainty than to list putative cluster candidates, re-examining the objects discovered in the past decades in the light of the \textit{Gaia}~DR2 catalogue in addition to the original discovery data appears to be a necessary task, which is greatly facilitated in the cases where the authors publish the individual list of stars they consider members of a potential cluster.

\section{Distribution in the Galactic disk} \label{sec:galaxy}
The distances inferred in Sect.~\ref{sec:distances} can be used to place the clusters on the Galactic plane. Their distribution is shown in Fig.~\ref{fig:map_XY_spiral_ages}. 
We notice that they clearly trace the Perseus arm, the local arm, and the Sagittarius arm of the Milky Way, and remark that the Perseus arm appears interrupted between $l=135^{\circ}$ and $l=240^{\circ}$.
Using the cluster ages listed in MWSC as a colour-code visually confirms that the youngest clusters are clearly associated with the spiral arms, while older clusters are distributed in a more dispersed fashion, which is naturally explained by the fact that spiral arms are the locus of star formation \citep[see e.g.][]{Dias05}. 
The lack of clusters tracing the Perseus arm is not due to a bias in our method, but to a general lack of known tracers in that direction, as already noted by \citet{Moitinho06} and \citet{Vazquez08}. 

\begin{figure*}[ht]
\begin{center} \resizebox{\hsize}{!}{\includegraphics[scale=0.5]{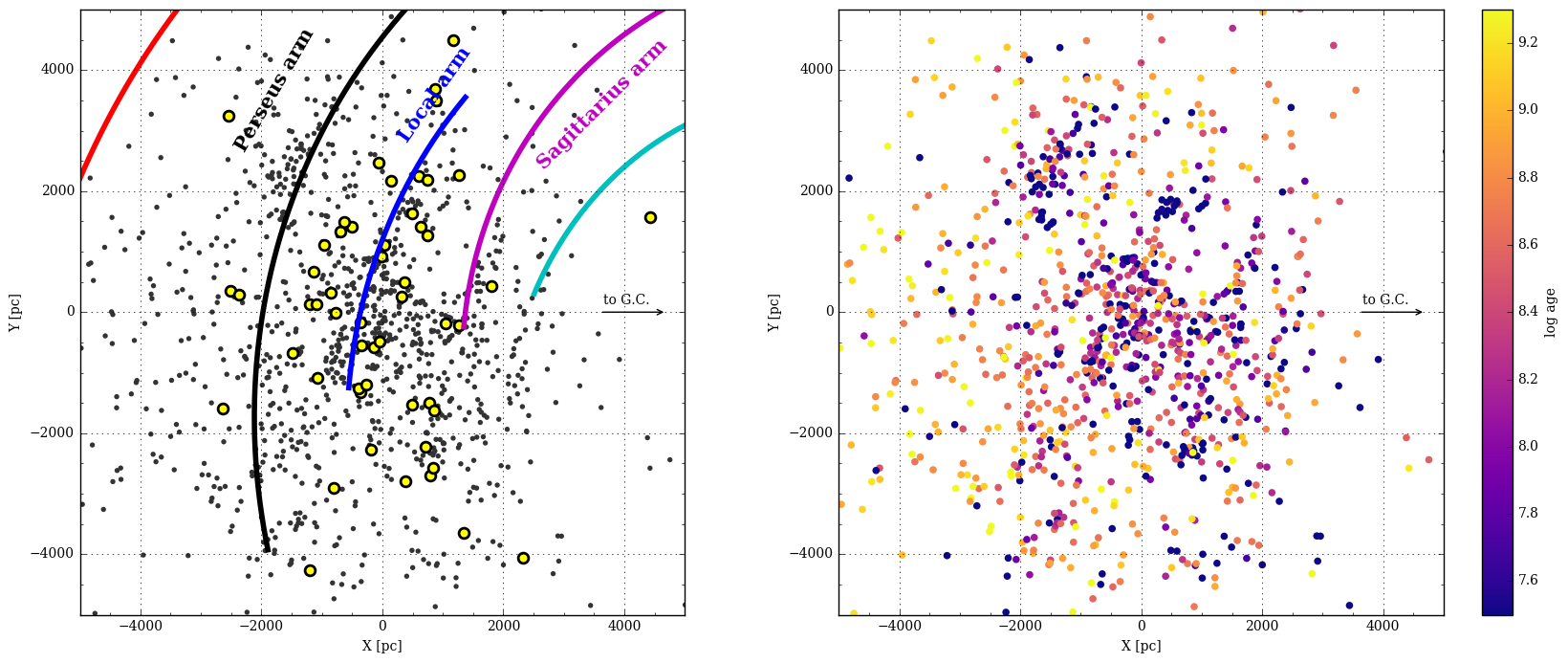}} \caption{\label{fig:map_XY_spiral_ages} Left: location of the OCs projected on the Galactic plane, using the distances derived in this study. The yellow dots indicate the objects newly identified in this study. Right: same sample of OCs, colour-coded by age (as listed in MWSC). Superimposed is the spiral arms model of \citet{Reid14}.} \end{center}
\end{figure*}

Figure~\ref{fig:map_Rgal_Z_ages_colorbar} shows that, as expected, young clusters are found near the plane, while older clusters can be found at all Galactic altitudes \citep[see e.g.][and references therein]{Lynga82,Bonatto06,Buckner14}. It is also apparent that fewer OCs are found at small Galactocentric radii, and that old clusters tend to be found in the outer disk. This observation was already made by \citet{Lynga82} \citep[see also e.g.][]{Lepine11}, and is here very obvious from Fig.~\ref{fig:map_Rgal_Z_ages_colorbar}. 

Although \citet{Becker70} consider that the spiral pattern is best traced with clusters younger than 55\,Myr, and \citet{Dias05} by clusters younger than 12\,Myr, we remark here that relatively older objects with ages in the range $100 - 300$\,Myr ($\log t \sim 8 - 8.5$) still seem to follow the spiral structure, which is in agreement with the studies of the Perseus arm of \citet{Moitinho06} and \citet{Vazquez08}
Given that the ages in literature and in catalogues show discrepant values for many clusters, this will be investigated in future works using \textit{Gaia} data and homogeneous analysis (Bossini et al., in prep.), as many ages (and in particular those of the clusters where our distances are at odds with MWSC) should be revised.

The median altitude Z for all clusters in this study is $-15.4\pm8.9$\,pc, $-15.3\pm5.2$\,pc when restricting the sample to those with estimated distances under 4\,kpc, and $-18.3\pm4.5$\,pc keeping only OCs younger than $\log t=8.5$. Those negative values correspond to a positive displacement of the Solar altitude (towards the north Galactic pole), in agreement with the results obtained by \citet{Bonatto06} ($14.8\pm2.4$\,pc) and \citet{Joshi07} ($17\pm3$\,pc) with open clusters. The running median traced in the bottom panel of Fig.~\ref{fig:map_Rgal_Z_ages_colorbar} shows local variations, in particular in the Galactocentric distance range $7.9 < R_{\mathrm{GC}} < 8.2$\,kpc (roughly corresponding to the location of the Local arm in the first quadrant) where the median altitude is $+22.0\pm14.5$\,pc. 

%

\begin{figure}[ht!]
\begin{center} \resizebox{\hsize}{!}{\includegraphics[scale=0.5]{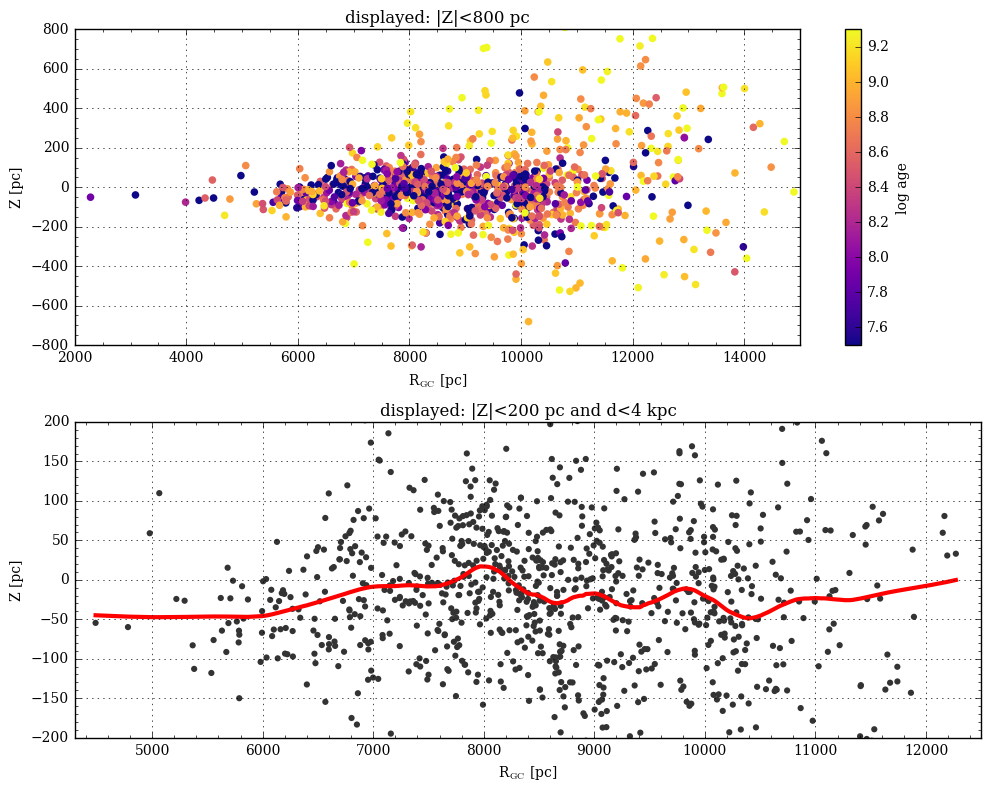}} \caption{\label{fig:map_Rgal_Z_ages_colorbar} Top: distance Z to the Galactic plane vs Galactocentric distance $R_{\mathrm{GC}}$, colour-coded by age (as listed in MWSC). Bottom: Z vs $R_{\mathrm{GC}}$ for OCs within 4\,kpc of the Sun. The red line is a lowess smoothing (local median based on the nearest 15\% of points in the sample). } \end{center}
\end{figure}

In Fig.~\ref{fig:position_withage} we display Z and $R_{\mathrm{GC}}$ as a function of age, showing that the median distance from the Galactic plane is about constant for clusters up to $\log t \sim 8.5$ (44\,pc), then increases with age (61\,pc for $8.5\log t<9$, 204\,pc for $\log t>9$). The apparent lack of young clusters with Galactocentric distances larger than 10\,kpc is mainly due to the gap in the Perseus arm.

\begin{figure}[ht]
\begin{center} \resizebox{\hsize}{!}{\includegraphics[scale=0.5]{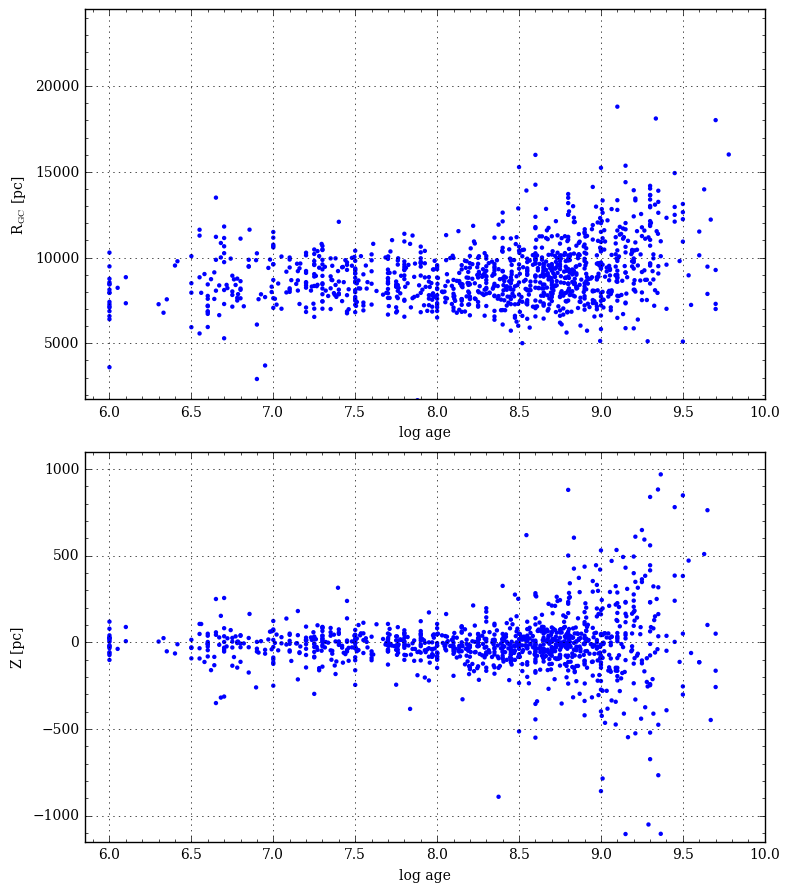}} \caption{\label{fig:position_withage} Top: Galactocentric distance (using the distances derived in this study) against age (from MWSC) for all OCs identified in this study. Bottom: distance Z from the plane against MWSC age for the same set of OCs.  } \end{center}
\end{figure}

\section{Discussion} \label{sec:discussion}

Although several authors have claimed (or assumed) that the census of OCs was complete out to distances of 1.8\,kpc, the recent discoveries of \citet{Castro18} who found 31 OCs in TGAS data, half of which closer than 500\,pc, as well as the serendipitous discoveries reported in this study, confirm the doubts expressed by \citet{Moitinho10} and reopen the question on how many clusters remain to be found with distances under a kiloparsec. A dedicated search, possibly combined with non-\textit{Gaia} data such as near-infrared photometry, radial velocities, and deep all-sky surveys would no doubt reveal more clusters in the vicinity of the Sun, and extend the cluster discoveries to larger distances. \textit{Gaia} data has also revealed new distant objects at high Galactic latitudes \citep{Koposov17,Torrealba18}, which can be further characterised through dedicated studies.

The main goal of this study is to list cluster members and derive astrometric parameters and distances from \textit{Gaia}~DR2 data alone. The precision and depth of the \textit{Gaia} photometry coupled with the ability to distinguish cluster members from their astrometry allows to determine reliable cluster ages for an unprecedently large sample of clusters. Accurate fitting of isochrones to photometric data however relies on assumptions on the chemical compositions of the stars under study (most importantly their metallicity and alpha-abundance) and can be much improved when the metallicity is accurately known \citep[see e.g.][]{Randich18}. The observational campaigns of the \textit{Gaia}-ESO Survey \citep{Gilmore12}, APOGEE \citep{Frinchaboy13}, or GALAH \citep{Martell17} have OC stars among their targets, but less than 5\% of the currently known OCs have been studied through means of high-resolution spectroscopy. The ambition of deriving cluster ages is beyond the scope of this paper, and will be treated in Bossini et al. (in prep.).

The results presented in Sect.~\ref{sec:galaxy} clearly show that old and young clusters present distinct distributions, with old clusters being found further from the Galactic plane. The striking absence of high-altitude clusters with $R_{\mathrm{GC}}<7$\,kpc could be due to clusters in the dense environment of the inner disk being disrupted before having the time to move to higher orbits. Possible scenarios invoking outwards migration could be tested if reliable metallicity determinations were able to link some of the old, high-altitude OCs to a birthplace in the inner disk. 

In addition to the proper motions, \textit{Gaia}~DR2 contains radial velocities for 7 million stars, which we did not exploit in this study. Those velocities, complemented with ground-based observations, allow to infer Galactic orbits for a large number of OCs and understand how the different populations behave kinematically. The topic of Galactic orbits is under study by Soubiran et al. (in prep.).

\section{Summary and conclusion} \label{sec:conclusion}
In this paper we rely on \textit{Gaia} data alone and apply an unsupervised membership assignment procedure to determine lists of cluster members. We provide the membership and mean parameters for a set of 1229 clusters, including 60 newly discovered objects and two globular clusters previously classified as OCs. We derive distances from the \textit{Gaia}~DR2 parallaxes, and show the distribution of identified OCs in the Galactic disk. We make use of ages listed in the literature in order to confirm that young and old clusters have significantly different distributions, with young objects following more tightly the spiral arms and plane of symmetry of the Galaxy, while older clusters are found more dispersed and at higher altitudes. They are also rarer in the inner regions of the disk.

Open clusters have been a popular choice of tracers of the properties of the Galactic disk for decades, partly because their distances can be estimated relatively easily by means of photometry. They also constitude valuable targets to study stellar astrophysics. In the \textit{Gaia} era of sub-milliarcsecond astrometry, OCs still constitute valuable tracers, because the mean parallax of a group of stars can be estimated to a greater precision than for individual sources. The positions we obtain reveal the structure of the disk in a radius of 4\,kpc around our location.
It is however difficult to estimate distances from \textit{Gaia}~DR2 parallaxes alone for stars more distant than 10\,kpc, and the distance listed in this study for the distant clusters could be improved by the use of photometric information, possibly combined with astrometry in a Bayesian approach. Regardless of our ability to determine reliable distances to them, the number of available tracers with distances larger than 5\,kpc is not sufficient to draw a portrait of the Milky Way out to large distances. We therefore emphasize the need for further observational, methodological and data-analysis studies oriented towards the discovery of new OCs in the most distant regions of the Milky Way.

\section*{Acknowledgements}

TCG acknowledges support from Juan de la Cierva -formaci\'on 2015 grant, MINECO (FEDER/UE).

This work has made use of data from the European Space Agency (ESA) mission \textit{Gaia} (www.cosmos.esa.int/gaia), processed by the \textit{Gaia} Data Processing and Analysis Consortium (DPAC, www.cosmos.esa.int/web/gaia/dpac/consortium). Funding for the DPAC has been provided by national institutions, in particular the institutions participating in the \textit{Gaia} Multilateral Agreement. This work was supported by the MINECO (Spanish Ministry of Economy) through grant ESP2016-80079-C2-1-R (MINECO/FEDER, UE) and
ESP2014-55996-C2-1-R (MINECO/FEDER, UE) and MDM-2014-0369 of ICCUB
(Unidad de Excelencia `Mar\'ia de Maeztu').
AB acknowledges support from Premiale 2015 MITiC (PI B. Garilli)

The preparation of this work has made extensive use of Topcat \citep{Taylor05}, and of NASA's Astrophysics Data System Bibliographic Services, as well as the open-source Python packages Astropy \citep{Astropy13}, numpy \citep{VanDerWalt11}, and scikit-learn \citep{scikit-learn}.
The figures in this paper were produced with Matplotlib \citep{Hunter07} and Healpy, a Python implementation of HEALPix \citep{Gorski05}. 
%


\bibliographystyle{aa} 
\linespread{1.5}                
\bibliography{biblio}

\clearpage
\appendix

\section{Maps and colour-magnitude diagrams for a few selected clusters}

\begin{figure}[ht]
\begin{center} \resizebox{\hsize}{!}{\includegraphics[scale=0.5]{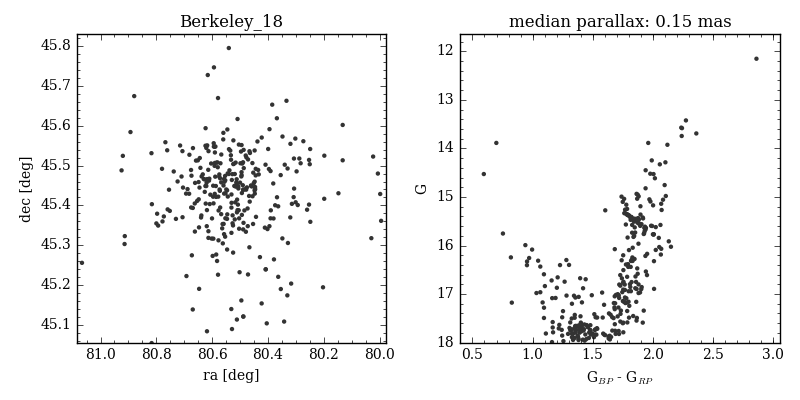}} \caption{\label{fig:example_Berkeley_18} Left: distribution of the probable members of Berkeley~18. Right: colour-magnitude diagram of the probable members.} \end{center}
\end{figure}

\begin{figure}[ht]
\begin{center} \resizebox{\hsize}{!}{\includegraphics[scale=0.5]{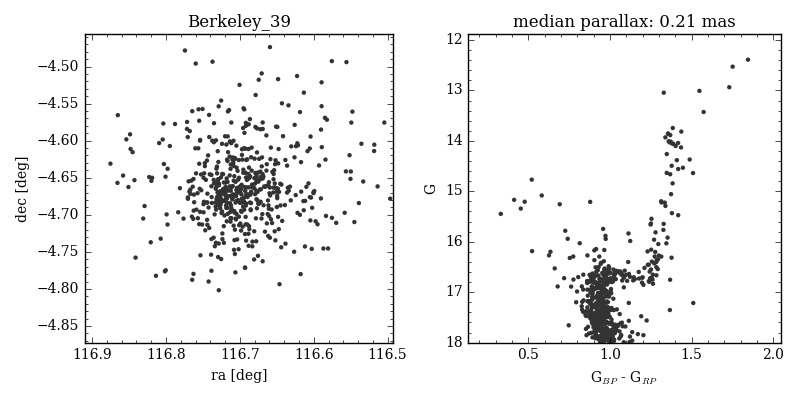}} \caption{\label{fig:example_Berkeley_39} Left: distribution of the probable members of Berkeley~39. Right: colour-magnitude diagram of the probable members.} \end{center}
\end{figure}

\begin{figure}[ht]
\begin{center} \resizebox{\hsize}{!}{\includegraphics[scale=0.5]{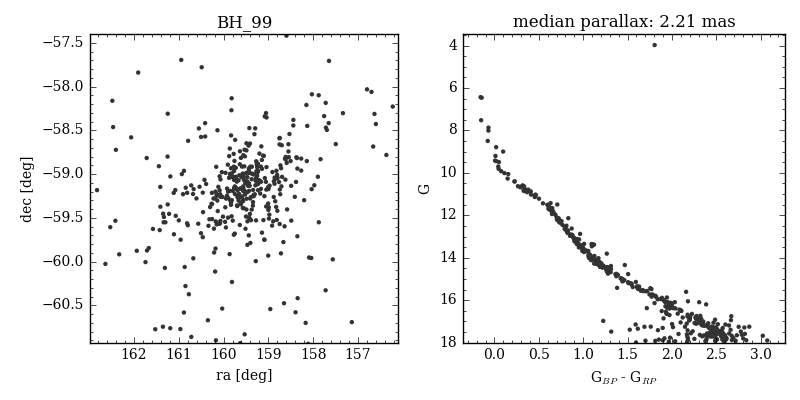}} \caption{\label{fig:example_BH_99} Left: distribution of the probable members of BH~99. Right: colour-magnitude diagram of the probable members.} \end{center}
\end{figure}

\begin{figure}[ht]
\begin{center} \resizebox{\hsize}{!}{\includegraphics[scale=0.5]{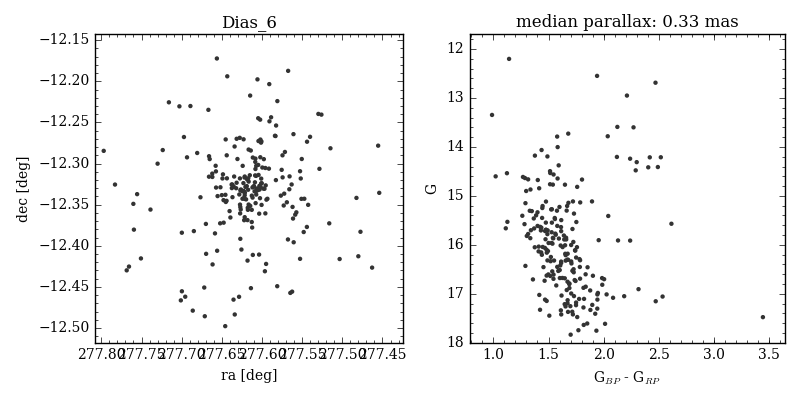}} \caption{\label{fig:example_Dias_6} Left: distribution of the probable members of Dias~6. Right: colour-magnitude diagram of the probable members.} \end{center}
\end{figure}

\begin{figure}[ht]
\begin{center} \resizebox{\hsize}{!}{\includegraphics[scale=0.5]{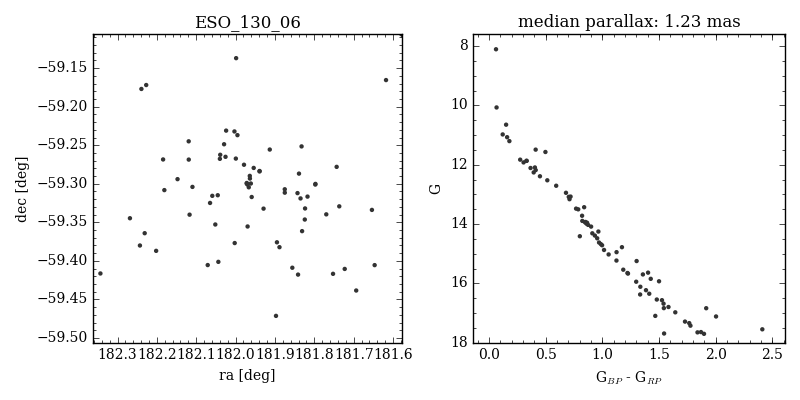}} \caption{\label{fig:example_ESO_130_06} Left: distribution of the probable members of ESO~130~06. Right: colour-magnitude diagram of the probable members.} \end{center}
\end{figure}

\begin{figure}[ht]
\begin{center} \resizebox{\hsize}{!}{\includegraphics[scale=0.5]{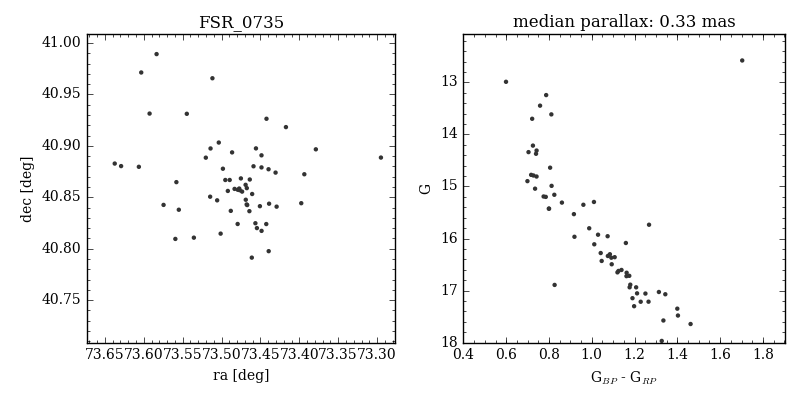}} \caption{\label{fig:example_FSR_0735} Left: distribution of the probable members of FSR~0735. Right: colour-magnitude diagram of the probable members.} \end{center}
\end{figure}

\begin{figure}[ht]
\begin{center} \resizebox{\hsize}{!}{\includegraphics[scale=0.5]{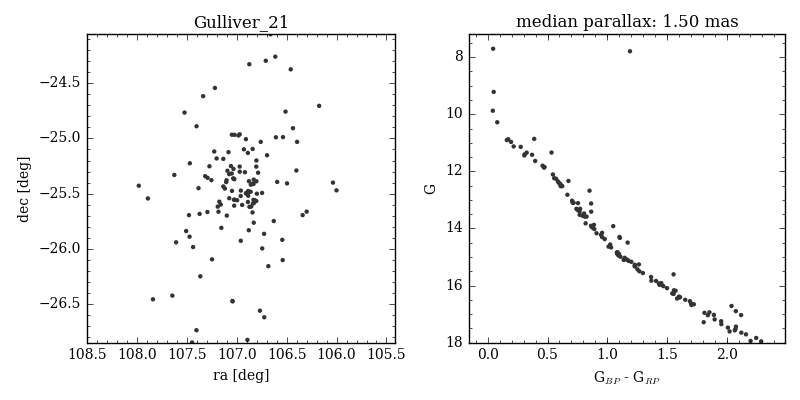}} \caption{\label{fig:example_Gulliver_21} Left: distribution of the probable members of Gulliver~21. Right: colour-magnitude diagram of the probable members.} \end{center}
\end{figure}

\clearpage

\begin{figure}[ht]
\begin{center} \resizebox{\hsize}{!}{\includegraphics[scale=0.5]{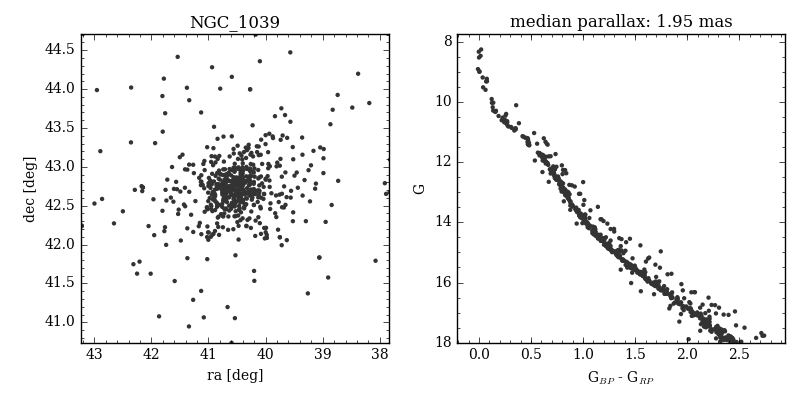}} \caption{\label{fig:example_NGC_1039} Left: distribution of the probable members of NGC~1039. Right: colour-magnitude diagram of the probable members.} \end{center}
\end{figure}

\begin{figure}[ht]
\begin{center} \resizebox{\hsize}{!}{\includegraphics[scale=0.5]{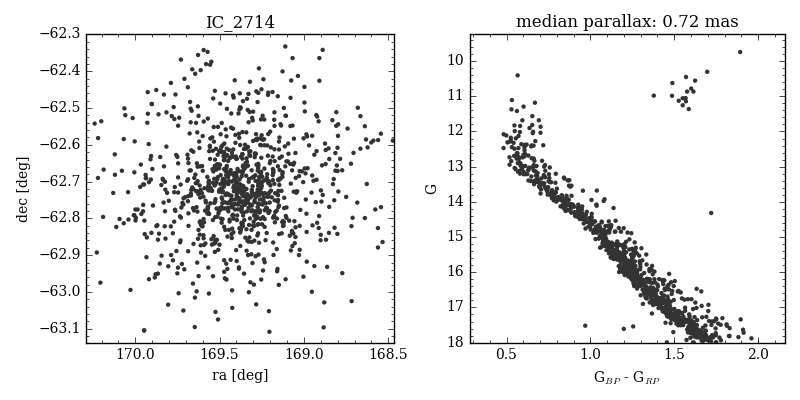}} \caption{\label{fig:example_IC_2714} Left: distribution of the probable members of IC~2714. Right: colour-magnitude diagram of the probable members.} \end{center}
\end{figure}

\begin{figure}[ht]
\begin{center} \resizebox{\hsize}{!}{\includegraphics[scale=0.5]{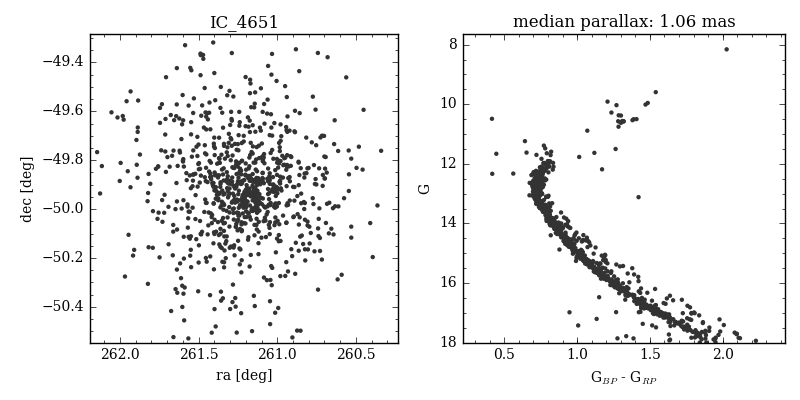}} \caption{\label{fig:example_IC_4651} Left: distribution of the probable members of IC~4651. Right: colour-magnitude diagram of the probable members.} \end{center}
\end{figure}

\begin{figure}[ht]
\begin{center} \resizebox{\hsize}{!}{\includegraphics[scale=0.5]{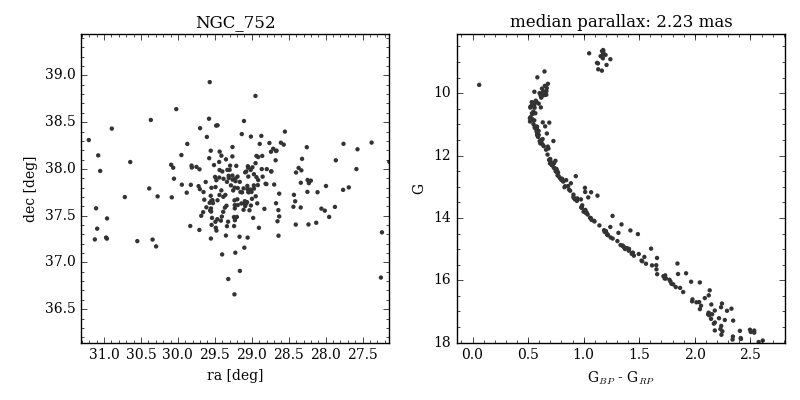}} \caption{\label{fig:example_NGC_752} Left: distribution of the probable members of NGC~752. Right: colour-magnitude diagram of the probable members.} \end{center}
\end{figure}

\begin{figure}[ht]
\begin{center} \resizebox{\hsize}{!}{\includegraphics[scale=0.5]{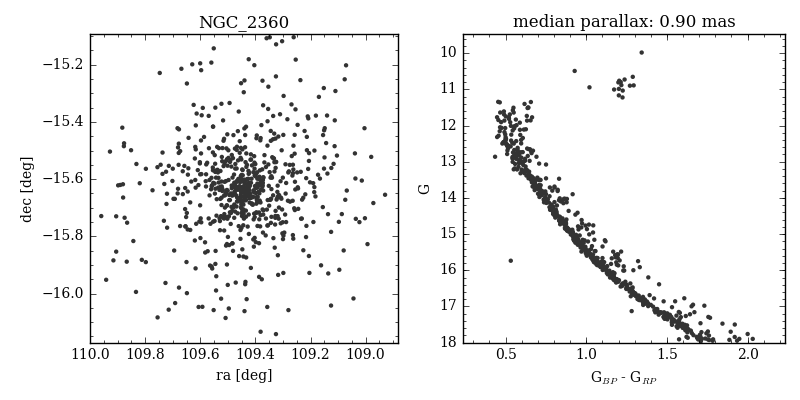}} \caption{\label{fig:example_NGC_2360} Left: distribution of the probable members of NGC~2360. Right: colour-magnitude diagram of the probable members.} \end{center}
\end{figure}

\begin{figure}[ht]
\begin{center} \resizebox{\hsize}{!}{\includegraphics[scale=0.5]{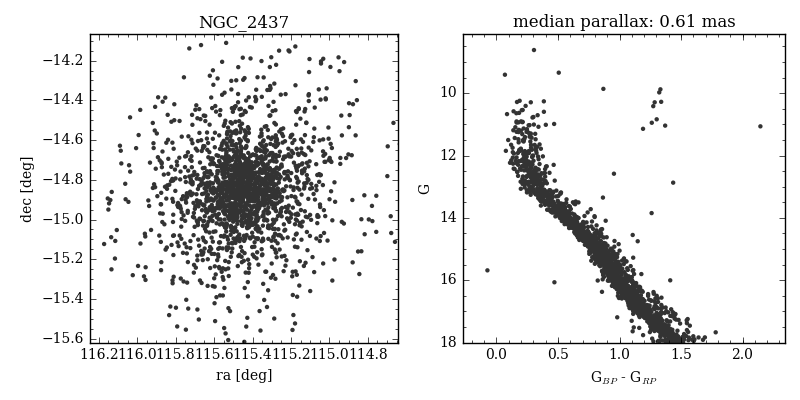}} \caption{\label{fig:example_NGC_2437} Left: distribution of the probable members of NGC~2437. Right: colour-magnitude diagram of the probable members.} \end{center}
\end{figure}

\begin{figure}[ht]
\begin{center} \resizebox{\hsize}{!}{\includegraphics[scale=0.5]{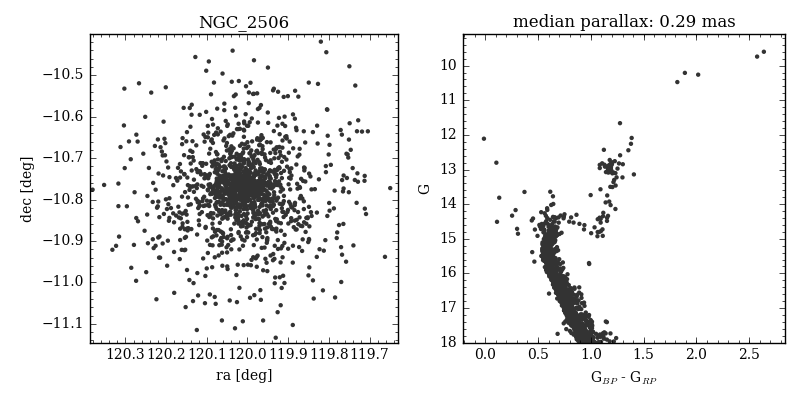}} \caption{\label{fig:example_NGC_2506} Left: distribution of the probable members of NGC~2506. Right: colour-magnitude diagram of the probable members.} \end{center}
\end{figure}

\begin{figure}[ht]
\begin{center} \resizebox{\hsize}{!}{\includegraphics[scale=0.5]{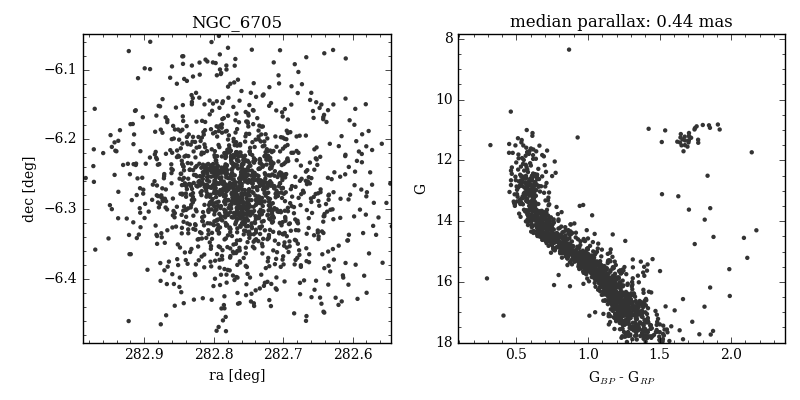}} \caption{\label{fig:example_NGC_6705} Left: distribution of the probable members of NGC~6705. Right: colour-magnitude diagram of the probable members.} \end{center}
\end{figure}

\begin{figure}[ht]
\begin{center} \resizebox{\hsize}{!}{\includegraphics[scale=0.5]{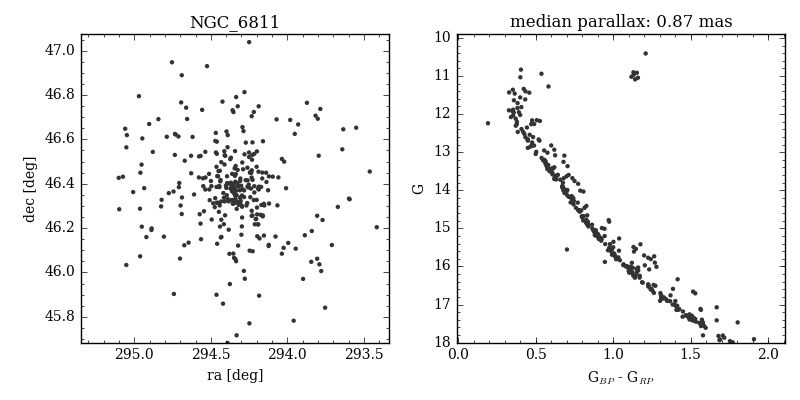}} \caption{\label{fig:example_NGC_6811} Left: distribution of the probable members of NGC~6811. Right: colour-magnitude diagram of the probable members.} \end{center}
\end{figure}

\begin{figure}[ht]
\begin{center} \resizebox{\hsize}{!}{\includegraphics[scale=0.5]{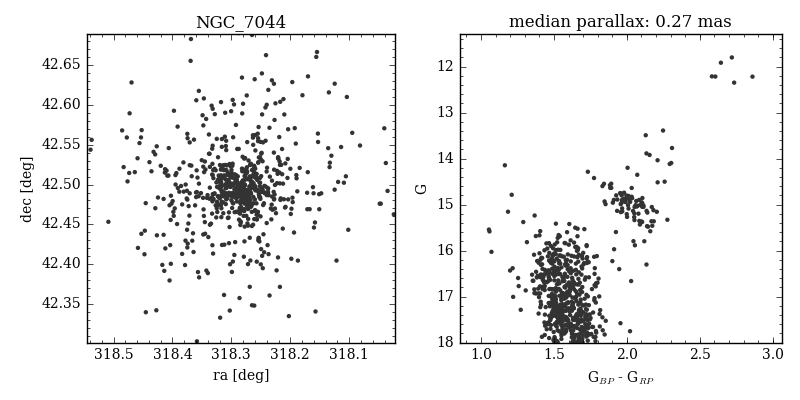}} \caption{\label{fig:example_NGC_7044} Left: distribution of the probable members of NGC~7044. Right: colour-magnitude diagram of the probable members.} \end{center}
\end{figure}

\begin{figure}[ht]
\begin{center} \resizebox{\hsize}{!}{\includegraphics[scale=0.5]{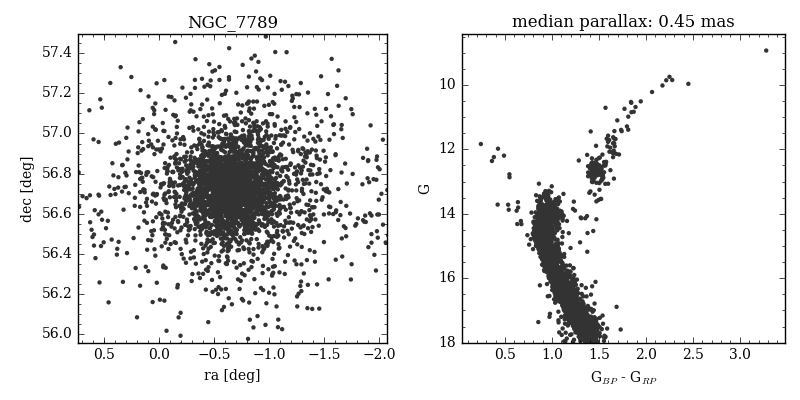}} \caption{\label{fig:example_NGC_7789} Left: distribution of the probable members of NGC~7789. Right: colour-magnitude diagram of the probable members.} \end{center}
\end{figure}

\begin{figure}[ht]
\begin{center} \resizebox{\hsize}{!}{\includegraphics[scale=0.5]{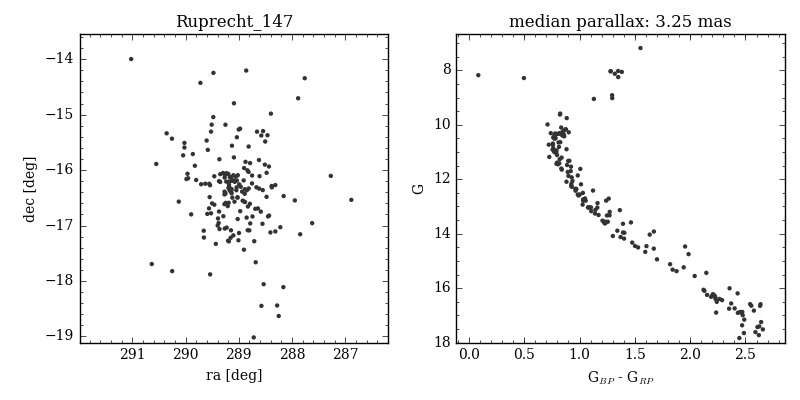}} \caption{\label{fig:example_Ruprecht_147} Left: distribution of the probable members of Ruprecht~147. Right: colour-magnitude diagram of the probable members.} \end{center}
\end{figure}

\begin{figure}[ht]
\begin{center} \resizebox{\hsize}{!}{\includegraphics[scale=0.5]{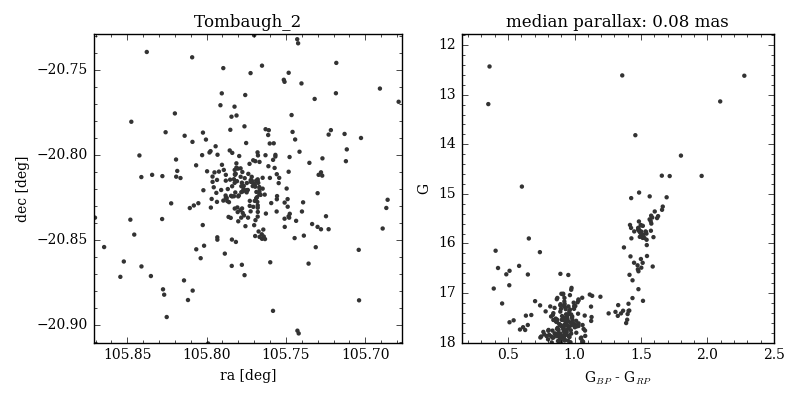}} \caption{\label{fig:example_Tombaugh_2} Left: distribution of the probable members of Tombaugh~2. Right: colour-magnitude diagram of the probable members.} \end{center}
\end{figure}

\end{document}